\documentclass[aip,jcp,12pt]{revtex4-1}

\usepackage{times} 
\usepackage{amssymb,amsmath}
\usepackage{graphics}
\usepackage{verbatim}
\usepackage[justification=raggedright]{caption,subfig}
\usepackage{natbib}
\usepackage{url}
\usepackage[usenames,dvipsnames]{color}
\usepackage{graphicx}
\usepackage[]{hyperref}
\hypersetup{
pdftitle={Parameter passing between Molecular Dynamics and continuum
  models for droplets on solid substrates: The static case} 
pdfauthor={N.Tretyakov, M.M\"uller, D.Todorova and U.Thiele},  
pdfproducer={lateX}, 
pdfview=FitV,       
pdfstartview=FitB,  
linkcolor=blue,     
citecolor=blue,     
urlcolor=blue,      
breaklinks=true,    
colorlinks=true,
citebordercolor=0 0 0,  
filebordercolor=0 0 0,
linkbordercolor=0 0 0,
menubordercolor=0 0 0,
urlbordercolor=0 0 0,
pdfhighlight=/I,
pdfborder=0 0 0,   
bookmarksopen=true,
bookmarksnumbered=true
}

\usepackage{ulem}

\newcommand{\ttuwe}[1]{\texttt{\textcolor{BrickRed}{\textbf{uwe:} #1}}}

\renewcommand{\vec}[1]{\mathbf{#1}}

\begin{document}
\title{Parameter passing between Molecular Dynamics and continuum
  models for droplets on solid substrates: The static case}
\author {Nikita Tretyakov}
\email{Nikita.Tretyakov@theorie.physik.uni-goettingen.de}
\author {Marcus M\"{u}ller}
\email{mmueller@theorie.physik.uni-goettingen.de}
\affiliation{Institut f\"{u}r Theoretische Physik, Friedrich-Hund-Platz 1,
37077 G\"{o}ttingen, Germany}
\author {Desislava Todorova}
\email{D.Todorova@lboro.ac.uk}
\author {Uwe Thiele}
\email{u.thiele@lboro.ac.uk}
\homepage{http://www.uwethiele.de}
\affiliation{Department of Mathematical Sciences, Loughborough University,
Loughborough, Leicestershire, LE11 3TU, UK}
\date{\today}

 \begin{abstract}
   We study equilibrium properties of polymer films and droplets on a
   solid substrate employing particle-based simulation techniques
   (Molecular Dynamics) and a continuum description. Parameter-passing
   techniques are explored that facilitate a detailed comparison of
   the two models. In particular, the liquid-vapor, solid-liquid
   and solid-vapor interface tensions, and the Derjaguin or
   disjoining pressure are determined by Molecular
   Dynamics simulations. This information is then introduced into continuum
   descriptions accounting for (i) the full curvature and (ii) a
   long-wave approximation of the curvature (thin film model). A
   comparison of the dependence of the contact angle on droplet size
   indicates that the theories agree well if the contact angles are
   defined in a compatible manner.
\end{abstract} 

\maketitle

\section{Introduction}
\label{sec:intro}

In the previous decade increasing attention has focused on the behavior of small quantities of liquid on hard \cite{Reit92,KoBa00,Beck03,QWS03,MR_SD_2008,BEIM2009,Thie10} or soft \cite{SC95,MM_CP_JS_2008} substrates in equilibrium or  under the influence of driving forces parallel to the substrate \cite{Thie01,JS_MM_2008,BM_HK_JY_2010}. Current research mainly considers two levels of description: particle-based models \cite{KoBa00,AM01,KPB02,LMcD_MM_KB_2002,MM2003,GG2003,CB2008,FL_JS_CP_MM_2011} and continuum theory \cite{BW1991,SD_MN_1991,MN_SD_1993,TVN01,Thie01,Beck03,DY2005,MR_SD_2008,FV_PT_BW_2008,BEIM2009,MG_FV_DR_2009,BM_HK_JY_2010,NM_FV_IS_2010}. The former describes the liquid in terms of the position and momenta of particles. These may represent atoms in a chemically realistic model or one lumps together a small number of atoms into an effective interaction center (called 'bead') in a coarse-grained model. 
The reduction of the number of degrees of freedom and the soft interactions in the coarse-grained description facilitate the study of long time and length scales. The properties of particle-based models are studied by discrete stochastic simulations, i.e., Monte-Carlo simulation or Molecular Dynamics. The advantage of retaining the particle degrees of freedom consists of the ability to refine the model towards a chemically realistic description and to include effects of thermal fluctuations and of discreteness of matter that are expected to become important on small length scales. However, these stochastic simulation techniques are limited to droplets of a linear size that does not exceed a few nanometers. 

Continuum models,
in turn, describe the liquid in terms of collective variables that do
not refer to individual particles. Typical examples of continuum
theories are the hydrodynamic description in terms of the density and
momentum fields or interface models that describe
the liquid only through a characterization of the motion of its
liquid-vapor boundary. Continuum descriptions can address engineering
time and length scales but depend on phenomenological material
  constants that are often not
related in a straightforward way to the microscopic interactions of
the particle-based description. Thus effort has to be devoted to
parameter-passing techniques that transfer information from
particle-based models to the continuum description. To this end, two
questions have to be addressed: (i) Which is the relevant information
of the particle-based model needed in the continuum description and
(ii) how can one extract this information from the particle-based
description in the appropriate continuum form?

In the present work, we use a coarse-grained particle model of a
polymer drop on a solid substrate, and a thin film description that
characterizes the droplet shape by the location, $h$, of the
liquid-vapor interface above the substrate. We explore the behavior of
small nano-drops where both descriptions are computationally
feasible. We extract the interface tensions and the Derjaguin or
disjoining pressure \cite{deGe85,MR_SD_2008,StVe09} from Molecular Dynamics
simulation of the particle-based model and pass them to continuum
model. Then both approaches are used to determine the equilibrium
contact angle of a droplet as a function of the size of the droplet
and of interaction strength between the liquid and the substrate.

  To our knowledge, such a parameter passing scheme has not yet
  been developped for the case of liquid droplets on solid
  substrates. However, the disjoining pressure itself can be extracted in grandcanonical ensemble~\cite{Muller2001,MM2003,VC_AW_2005,LGM2006,JRE2010}. Additionally, related works exist for other geometries in canonical ensemble, such as free standing films or films adsorbed in pores~\cite{HeHe10}.

  Bhatt \textit{et al.}~\cite{BNR02} extract a disjoining pressure as a
  function of chemical potential from MD simulations for a free standing film
  of a volatile Lenard-Jones liquid and compare the results with the
  ones of density functional theory. 
  Their
  approach consists in the definition of the disjoining pressure as the difference of
  normal pressure in the film and the pressure in the homogeneous
  liquid at the same chemical potential as the film. However, as discussed in
  section~\ref{sec:pass}, the measurement of the chemical potential in a canonical ensemble is difficult and requires additional simulations. Moreover, despite of truncated potentials, they relate the disjoining pressure with solely long-range van der Waals dispersion forces and provide therefore comparison to Hamaker theory. The short-range forces stay outside the scope of their research.

  A planar liquid film bounded by a solid and vapor is studied
  by Han \cite{Han08} using grandcanonical MD simulations with a truncated and shifted
  Lennard-Jones interaction. The disjoining pressure is extracted in
  a similar way as in Ref.~\cite{BNR02} and again is associated with only long-range dispersion forces.

Note that parameter passing
  from MD simulations to continuum hydrodynamics is also frequently
  done in the context of liquid flow close to solid substrates
  \cite{KoBa95,Hadj99,CKB01,QWS03,QWS04,PDT05}. However, as these works do either not
  involve free interfaces \cite{CKB01,PDT05} or do not extract the
  disjoining pressure \cite{Hadj99,QWS03,QWS04}, we do here not discuss them
  further.

Our manuscript is structured as follows. In section~\ref{sec:mod} we present the particle-based and continuum approaches.  Then, section~\ref{sec:pass} details how we pass the parameters from the particle-based model into the continuum description. The subsequent section~\ref{sec:res} presents the dependence of the equilibrium contact angle on droplet size for various interaction energies between the liquid and the substrate. In passing, we describe several ways to define the equilibrium contact angle and discuss their relation to the macroscopic Young-Laplace law. Section~\ref{sec:conc} concludes
and gives an outlook beyond the case of equilibrium droplets.

\section{Models}
\label{sec:mod}

\subsection{Molecular Dynamics (MD)}
\label{sec:md}

\begin{figure}[t]
\includegraphics[width=1.0\hsize]{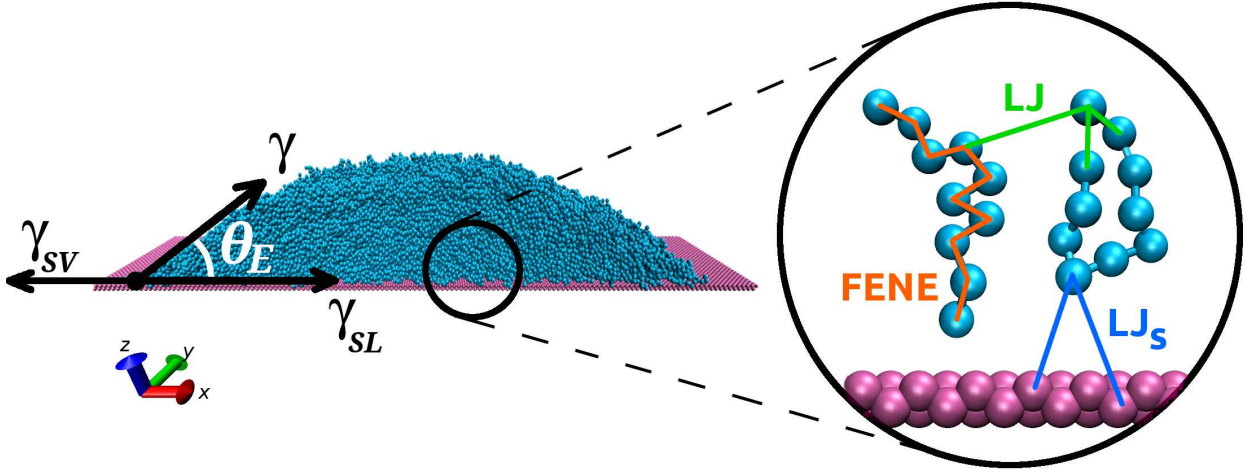}
\caption{Snapshot from MD simulation of a cylindrical drop with illustration of Young's equation (left). The enlargement close to the substrate (right) sketches the pairwise bead potentials. Coarse-grained beads of polymer chains (blue) interact with each other and with the substrate  modeled by two layers of face-\-centered-\-cubic lattice (lila)}
\label{fig:typical_int}
\end{figure}

Here, the mesoscopic discrete stochastic description is provided by Molecular Dynamics simulations of a widely used coarse-grained polymer model \cite{GG_KK_1986}, i.e., a polymer chain is not represented by each and every individual atom but it is modeled as a flexible, linear string of small conglomerates of atoms. These conglomerates are called ``beads''. The length of all polymer chains is fixed to $N_p = 10$ monomers~\cite{CP_KB_TK_MM_2006, JS_MM_2008} in all simulations. The potentials used in MD are represented in Fig.~\ref{fig:typical_int}

All bonded and non-bonded beads have unit mass, $m=1$, and interact via truncated and shifted Lennard-Jones (LJ) potentials
\begin{equation}
 U(r)=U_\textrm{{LJ}}(r)-U_\textrm{{LJ}}(r_c) 
\end{equation}
with
\begin{equation}
	U_\textrm{{LJ}}(r)=4 \epsilon \bigg[ \Big( \frac{\sigma}{r}
        \Big)^{12}-\Big( \frac{\sigma}{r} \Big)^6 \bigg]
\label{eq:lj}
\end{equation} 
if their distance is smaller than the cutoff distance $r_c = 2 \times 2^{1/6} \sigma$. $U_\textrm{{LJ}}(r_c)$ is the LJ potential evaluated at the cutoff distance. All LJ parameters are set to unity, $\epsilon = 1$ and $\sigma = 1$, i.e., we express all energies and lengths in units of $\epsilon$ and $\sigma$, respectively. The reduced time unit $\tau$ is set by a combination of the LJ parameters as $\tau = \sigma \sqrt{\frac{m}{\epsilon}}$.

The individual beads are connected into chains employing a finite extensible nonlinear elastic (FENE) potential given by \cite{Bird77, KK_GG_1990}
	\begin{equation}
	U_\textrm{FENE}=\left\{
	 \begin{array}{c c}
	 -\frac{1}{2}kR_{0}^2\:\ln\bigg[1-\Big(\frac{r}{R_{0}}\Big)^2\bigg] & \quad \text{for $r < R_{0}$}\\
	 \infty & \quad \text{for $r\geq R_{0}$}\\
	 \end{array} \right.
	\end{equation}
where $R_0 = 1.5 \sigma$ and $k = 30 \epsilon / \sigma^2$.

To control the wettability of the polymeric liquid on the substrate we account for the interaction of the beads with the solid substrate.
The substrate is modeled by a fixed array of atoms as in Ref.~\cite{JS_MM_2008} and not by an ideally smooth and homogeneous wall \cite{LMcD_MM_KB_2002, CP_KB_TK_MM_2006}. Specifically, the substrate is represented by two layers of a face-centered-cubic lattice of atoms with a number density of $\rho_\textrm{s} = 1.5 \sigma ^{-3}$. We also employ a truncated and shifted LJ interaction between the beads of the liquid and the individual constituents of the substrate
\begin{equation}
 U^\textrm{s}(r)=U^\textrm{s}_\textrm{{LJ}}(r)-U^\textrm{s}_\textrm{{LJ}}(r_c) 
\end{equation}
with
the length scale $\sigma_\textrm{s} = 0.75 \sigma$. The strength of interaction $\epsilon_\textrm{s}$ is varied. By changing $\epsilon_\textrm{s}$ from $0.2\epsilon$ to $\epsilon$, one tunes the wettability of the system from non-wetting (polymer droplet with a contact angle of $\theta_\mathrm{E}=163^\mathrm{o}$) to complete wetting (polymer film with $\theta_\mathrm{E}=0^\mathrm{o}$). 

All simulations are carried out in a computational domain that corresponds to a three dimensional box. Periodic boundary conditions are used in the $x$- and $y$-directions, whereas the range in the $z$-direction is limited by a repulsive ideal wall that is positioned far above the polymer liquid. The domain side lengths, $L_x$ and $L_y$, are chosen in such a way that one may study polymer films, $L_x = L_y$), and  two-dimensional drops (i.e., ridges in 3d), $L_y \ll L_x$. These ridges span the simulation box in $y$ direction and have the cylindrical form whose cross-section is well visible in Fig.~\ref{fig:typical_int}. $L_y$ is limited by the Plateau-Rayleigh instability that results in the instability of liquid ridges above a critical length. However, as this instability is normally subcritical \cite{BKHT11}, in a MD simulation $L_y$ has to be smaller than a critical ridge length $L_\mathrm{nl}$ that is smaller than the one resulting from the linear stability analysis of a ridge.

The radius of a 2d drop (3d ridge) scales as $\sqrt{N}$ (in comparison to $N^{1/3}$ for a spherical 3d drop), allowing us to study larger droplets \cite{JS_MM_2008}. Moreover, the length of the three-phase contact line, $2L_{y}$, is independent of the 2d droplet size. Thus, there is no direct effect of the line tension on the shape of the droplet.

The temperature of the system is controlled by a dissipative particle dynamics (DPD) thermostat \cite{PH_JK_1992,PE_PW_1995}. In DPD, the total force on a given monomer is given by
\begin{equation}
F_\mathrm{tot} = \sum_{j \neq i}(\mathbf{F}_{ij}+\mathbf{F}_{ij}^\mathrm{D}+\mathbf{F}_{ij}^\mathrm{R}),
\end{equation}
where the conservative force $\mathbf{F}_{ij}$ is derived from the potential between monomer $i$ and monomer $j$, $\mathbf{F}_{ij}^\mathrm{D}$ is a dissipative force and $\mathbf{F}_{ij}^\mathrm{R}$ is a random force. The dissipative and random forces act on pairs of particles and are of the form
\begin{equation}
 \mathbf{F}_{ij}^\mathrm{D}=-\gamma_\textrm{DPD}\; \omega_\textrm{D}(r_{ij})(\vec{e_{ij}} \cdot \vec{v_{ij}}) \vec{e_{ij}},
\end{equation}
\begin{equation}
 \mathbf{F}_{ij}^\mathrm{R}= \zeta\; \omega_\textrm{R}(r_{ij}) \theta_{ij} \vec{e_{ij}},
 \label{eq:dpd_rand}
\end{equation}
where $r_{ij}=\lvert \vec{r_i}-\vec{r_j} \rvert$, and the unit vector $\vec{e_{ij}} = \vec{r_{ij}} / r_{ij}$ points from the $j-$th to the $i-$th particle. In order to obey the fluctuation-dissipation theorem, the damping coefficient, $\gamma_\textrm{DPD}$, is connected to the amplitude of the noise, $\zeta$, via the fluctuation-dissipation theorem $\zeta^{2}=2 k_\textrm{B} T \gamma_\textrm{DPD}$ and the weight functions are defined as
\begin{equation}
 \omega_\textrm{R}^2(r_{ij}) = \omega_\textrm{D}(r_{ij}) = \left\{
	 \begin{array}{c c}
	 (1-\frac{r_{ij}}{r_c})^2 & \quad \text{for $r < r_c$}\\
	 0 & \quad \text{for $r\geq r_c$}\\
	 \end{array} \right.
\end{equation}
We fix $\gamma_\textrm{DPD}=0.5$ in all our simulations. The term $\theta_{ij}$ in Eq.~(\ref{eq:dpd_rand}) is a random noise term such that $\theta_{ij}=\theta_{ji}$ and its first and second moments are
\begin{equation}
 \langle \theta_{ij} \rangle = 0,
\end{equation}
\begin{equation}
 \langle \theta_{ij}(t)\theta_{kl}(t') \rangle = (\delta_{ik}\delta_{jl}+\delta_{il}\delta_{jk}) \delta(t-t').
\end{equation}
We use uniformly distributed random numbers  \cite{BD_WP_91} with the first and second moments dictated by the relations above.

Since the dissipative and random forces satisfy Newton's third law, they locally conserve momentum, i.e., they preserve the hydrodynamics of the flow (in contrast to the dissipative macroscopic behavior in Brownian dynamics). Using this DPD thermostat, we maintain the constant temperature, $k_\mathrm{B} T = 1.2 \epsilon$.  The equations of motion are integrated with the velocity Verlet algorithm \cite{WS_HA_PB_KW_1982} with a time step $\Delta t = 0.005 \tau$. We performed the simulations on GPU facilities using the HOOMD Software~\cite{HOOMD_site, JA_CL_AT_2008, CP_JA_SG_2011}.

The MD simulations are used to determine parameters that are passed on to the continuum model. Before the parameter passing is described in section~\ref{sec:pass}, we introduce in the following section the continuum model.

\subsection{Continuum model (CM)}
\label{sec:tfm}

We employ a highly coarse-grained description to characterize the free-energy of a droplet on a planar substrate in terms of the position of the solid-liquid and liquid-vapor interfaces. Generally, the free energy takes the translationally and rotationally invariant form
\begin{equation}
F = \gamma_\mathrm{SL} \int_\mathrm{SL} \mathrm{d}S + \gamma \int_\mathrm{LV} {\rm
  d}S +  \int_\mathrm{LV} \mathrm{d}S  \int_\mathrm{SL} \mathrm{d}S' \;
\tilde{g}(|\mathbf{r}-\mathbf{r}'|)
\label{eq:ff}
\end{equation}
where the integrals extend over the solid-liquid (SL) and liquid-vapor (LV) interfaces \footnote{If there exist 			additional long-ranged interactions, $V_\mathrm{lr}$,
	between the liquid and the solid, then one has the additional contribution
	$F_\mathrm{lr} = \int_\mathrm{L} \mathrm{d}^{3}\mathbf{r}\int_\mathrm{S}
	\mathrm{d}^{3}\mathbf{r}'\; V_\mathrm{lr}(|\mathbf{r}-\mathbf{r}'|)$.
	Writing $V_\mathrm{lr}(|\mathbf{r}|) = \nabla \cdot \mathbf{r}\Phi_\mathrm{lr}(|\mathbf{r}|)$,
	we obtain for the long-range contribution $F_\mathrm{lr} = \int_{\mathrm{SL}
	\cup \mathrm{LV}} \mathrm{d}\mathbf{S} \cdot \int_\mathrm{S} \mathrm{d}^{3}\mathbf{r}'\;
	(\mathbf{r}-\mathbf{r}') \Phi_\mathrm{lr}(|\mathbf{r}-\mathbf{r}'|)
	= \int_{\mathrm{SL} \cup \mathrm{LV}} \mathrm{d}\mathbf{S} \cdot
	\mathbf{e}_{x} g_\mathrm{lr}(h)$ with $g_\mathrm{lr}(h) = \int_\mathrm{S}
	\mathrm{d}^{3}\mathbf{r}'\; \mathbf{e}_{x}\cdot(\mathbf{r}-\mathbf{r}')
	\Phi_\mathrm{lr}(|\mathbf{r}-\mathbf{r}'|) = L_{y}\int \mathrm{d}x\;
	g_\mathrm{lr}(h) + \mathrm{const}$.}.
In Eq.~(\ref{eq:ff}), $\gamma_\mathrm{SL}$ and $\gamma$ are the solid-liquid and liquid-vapor interface tensions, respectively. The last term of Eq.~(\ref{eq:ff}) describes the effective interaction between the interfaces, and $\mathbf{r}$ and $\mathbf{r}'$ are points on the liquid-vapor and solid-liquid interface, respectively. In the following, we restrict our attention to 2d droplets on a planar substrate (cf.~Fig.~\ref{fig:typical_int}), choose the $x$-coordinate along the planar solid substrate and denote by $z=h(x)$ the local distance between a point $\mathbf{r} \equiv (x,y,z=h(x))$ of the liquid-vapor interface and the planar substrate (Monge representation). The interaction of a point on the liquid-vapor interface with the solid is obtained by integrating over the substrate area
\begin{equation}
g(h) = \int_\mathrm{SL} \mathrm{d}S'\; \tilde{g}(|\mathbf{r}-\mathbf{r}'|)
\end{equation}
which for a homogeneous substrate only depends on the distance, $h$, due to symmetry. $g(h)$ is the effective integrated interaction between a point of the liquid-vapor interface with the homogeneous, planar substrate, and it is termed interface potential. In this special case, the free energy functional (\ref{eq:ff}) takes the form
\begin{equation}
F[h] =   \gamma_\textrm{SL}  L_{y} \int \mathrm{d}x \; + L_{y} \int \mathrm{d}x \sqrt{1 + \left(
\partial_{x}h\right)^{2}} \; \Big[ \gamma + g(h) \Big],
\label{eqn:full}
\end{equation}
where $L_{y}$ denotes the system dimension parallel to the cylinder axis. In the limit that the equilibrium contact angle is small, one can adopt a long-wave approximation (or small-gradient expansion)
\begin{equation}
F[h] \approx \gamma_\textrm{SL}  L_{y} \int \mathrm{d}x \;  + L_{y} \int \mathrm{d}x \; \Big[1 + \frac{1}{2} \left(
\partial_{x}h \right)^{2} + \cdots \Big] \; \Big[ \gamma + g(h) \Big] ,
\label{eq:full_approx}
\end{equation}
It is important to note that, away from the droplet, there is a thin film of thickness $h_\textrm{min}$ with a flat liquid-vapor interface (dewetted surface). $h_\textrm{min}$ corresponds to the minimum of the interface potential. Eq.~(\ref{eq:full_approx}) yields for this dewetted part of the surface
\begin{equation}
 F[h_\textrm{min}] = L_{y} \int_{\substack{\textrm{dew} \\ \textrm{surf}}} \mathrm{d}x \, \Big[\gamma_\textrm{SL} + \gamma + g(h_\textrm{min})\Big] \equiv L_{y} \int_{\substack{\textrm{dew} \\ \textrm{surf}}} \mathrm{d}x \, \gamma_\textrm{SV} \, .
\end{equation}
Here, $\gamma_\textrm{SV} = \gamma_\textrm{SL} + \gamma + g(h_\textrm{min})$ is the solid-vapor interface tension. We emphasize that it is not a solid-vacuum surface free energy per unit area, $F_0$, that is half of the work needed to cut the bonds of a solid of a unit cross section into two equal pieces in vacuum. Moreover, as long as the solid is not altered by the contact with the liquid or vapor, its free energy per unit area remains constant, $F_0=\textit{const}$, and serves as the reference point for solid-liquid and liquid-vapor interface tensions. In our model of the solid we do not consider interactions between its constituents. Therefore, the work needed to cut the solid is zero and the reference value of the surface free energy per unit area is $F_0=0$.

The equilibrium shape of the droplet is obtained by minimizing this free energy functional subject to the constraint of fixed droplet volume
\begin{equation}
V_\mathrm{drop} =  L_{y} \int \mathrm{d}x \; h(x) = \mathrm{const}
\label{eq:constraint}
\end{equation}
yielding the condition
\begin{equation}
\pi(x) = -\frac{1}{L_{y}}\frac{\delta F}{\delta h(x)}  = \lambda \label{eq:el}
\end{equation}
where $\lambda$ is a Lagrange multiplier constraining the droplet volume. Using Eq.~(\ref{eqn:full}) we obtain
\begin{eqnarray}
\pi(x) &=& -\sqrt{1+\left(\partial_{x}h\right)^{2}}\Big[ \partial_{h}g \Big] + \partial_{x} \left(  \frac{\partial_{x}h}{\sqrt{1+\left(\partial_{x}h\right)^{2}}} \Big[ \gamma + g(h) \Big]\right) \nonumber\\
&=& \frac{\partial_{xx}h  \Big[ \gamma + g(h) \Big]}{\left[
    1+\left(\partial_{x}h\right)^{2} \right]^{3/2}}
-\frac{\partial_{h}g}{\sqrt{1 + \left(\partial_{x}h\right)^{2}}}
\label{eq:pi-full}
\end{eqnarray}
In the limit of small contact angles, $|\partial_{x}h|\ll1$, this equation adopts the form
\begin{equation}
\pi(x) = \partial_{xx}h  \Big[ \gamma + g(h) \Big] - \partial_{h}g.
\label{eq:pi-longwave}
\end{equation}
The pressure~(\ref{eq:pi-full}) consists of two contributions: (i) the curvature pressure, where $\kappa_\mathrm{full}=\frac{\partial_{xx}h}{\left[ 1+\left(\partial_{x}h\right)^{2}  \right]^{3/2}}$ is the curvature and $\gamma+g(h)$ is the effective tension of the interface a distance $h$ away from the solid substrate and (ii) the Derjaguin (or disjoining) pressure $\Pi(h)=- \partial_h g(h)$ that models wettability \cite{deGe85, StVe09}. The dimensionless ratio $g(h)/\gamma$ dictates the shape of a drop in the continuum model and it is this parameter that we will extract from the particle-based model in Sec.~\ref{sec:pass}.

A spatially non-uniform pressure, $\pi(x)$, gives rise to a flow of liquid inside the film. Using the Navier--Stokes equation and employing the long-wave approximation \cite{ODB97,Thie07,Thie10}, one
obtains
\begin{equation}
\partial _t h = -\partial_x \Gamma = -\partial _x \{ Q(h)\partial_x \pi(x) \} .
\label{eq:tfe}
\end{equation}
Here $Q(h) = h^3/3\eta$ is the mobility, $\eta$ is the dynamic viscosity of the liquid.  Note, that $\Gamma$ is a flux that is written as the product of a mobility and a pressure gradient. Eq.~(\ref{eq:tfe}) with (\ref{eq:pi-longwave}) is sometimes called a thin-film or lubrication model.

\begin{figure}[t]
{\large (a)}
\includegraphics[width=0.45\hsize]{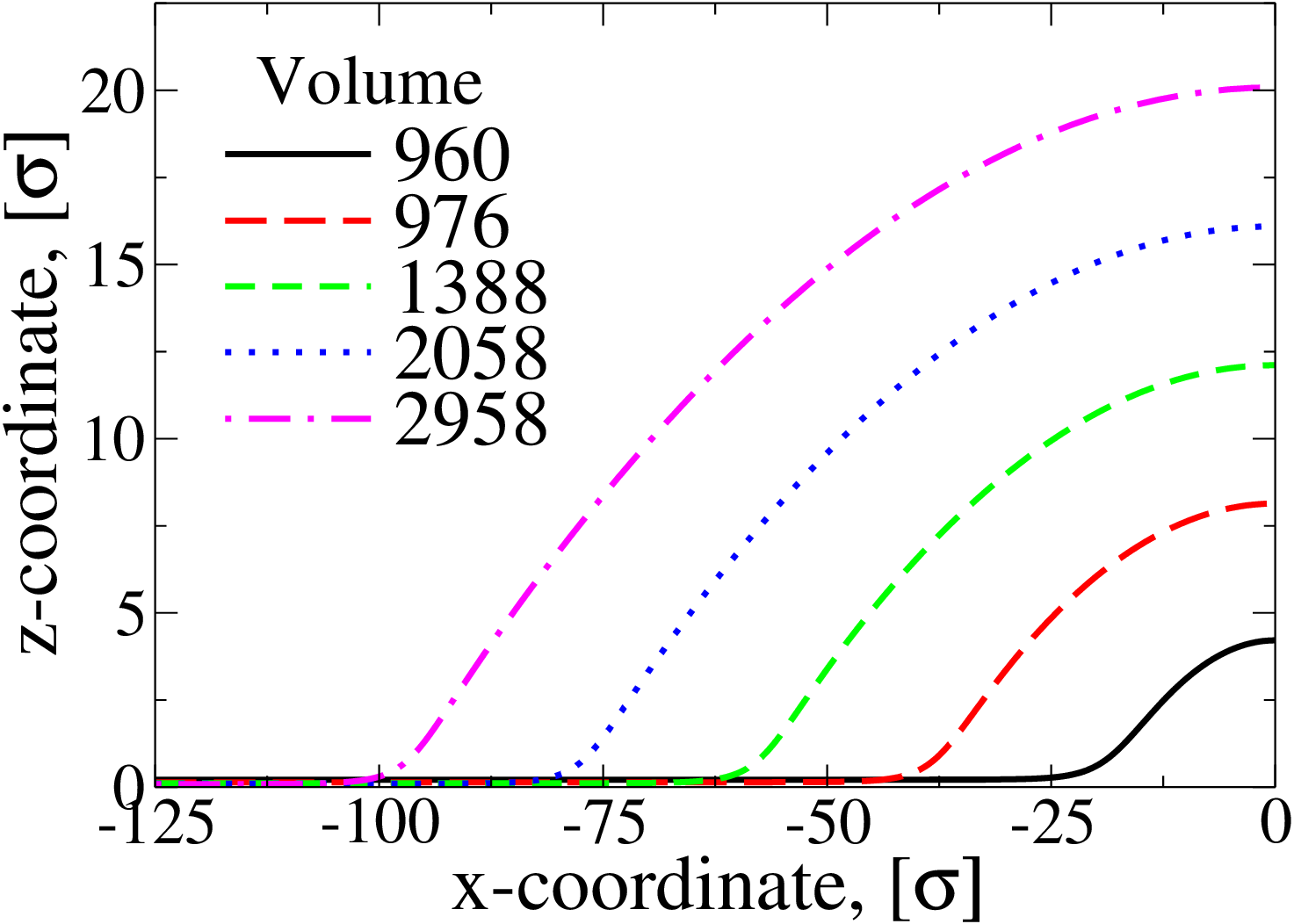}
\includegraphics[width=0.45\hsize]{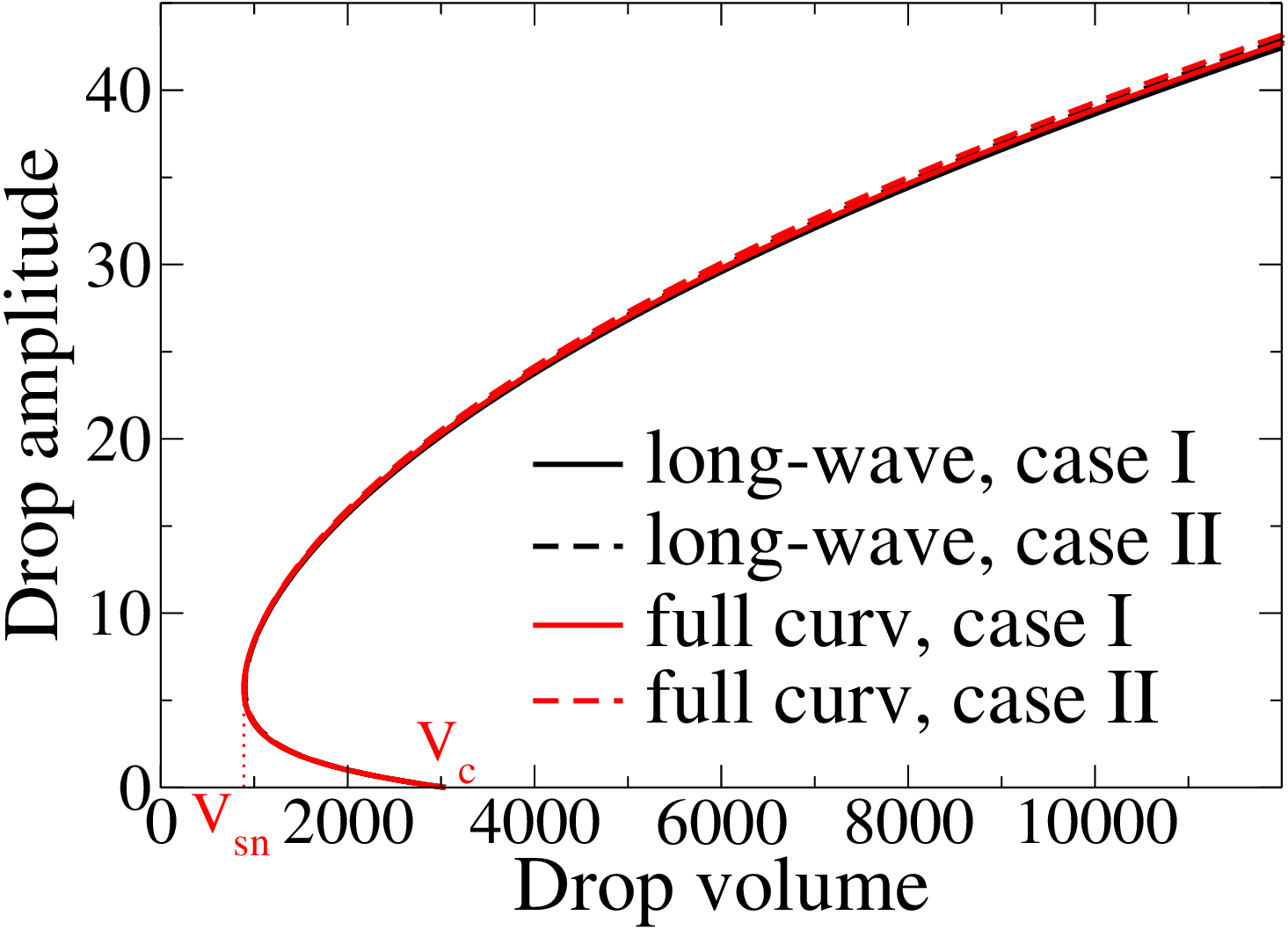}{\large (b)}
\caption{(a) Shown are selected half-profiles of droplets at volumes as given in the
  legend and (b) the bifurcation diagram presenting
  the drop height in dependence of the drop volume.
Calculations are performed
  with (i) the full curvature, i.e., Eq.~(\ref{eq:el}) with
  (\ref{eq:pi-full}), and (ii) the long-wave curvature, i.e., Eq.~(\ref{eq:el}) with
  (\ref{eq:pi-longwave}). Case I and II refer to usage of only $\gamma$
  or the full $\gamma+g(h)$ as prefactor of curvature,
  respectively. The profiles in panel (a) are obtained with case I for
  full curvature. The
  volume is controlled through appropriately adapting the Lagrange
  multiplier $\lambda$ at fixed domain size $L=4000$.  The employed
  disjoining pressure and interface tensions are extracted from MD
  simulations at $\epsilon_\mathrm{s}=0.81\epsilon$ (equivalent to 
an equilibrium contact angle of $\theta_\textrm{E}=23.57^o$, for details see
  below section~\ref{sec:res}).
}
\label{fig:typical-int}
\end{figure}

The equation describing stationary solutions may either be obtained by directly minimizing the functional $F[h]$ according to Eq.~(\ref{eq:el}) or, alternatively, one sets $\partial_t h=0$ in Eq.~(\ref{eq:tfe}) and integrates twice taking into account that $\Gamma=0$ in the steady state.  Here we use numerical continuation techniques \cite{DKK91} to solve the resulting ordinary differential equation as a boundary value problem on a domain of size $L$ with boundary conditions such that the center of the resulting drop solution is positioned on the right boundary $(x = 0)$ and on the left boundary $(x=-L)$ the profile approaches a precursor film. The volume is controlled by the integral condition, Eq.~\ref{eq:constraint}. Figure~\ref{fig:typical-int}(a) presents typical drop profiles for various volumes whereas Fig.~\ref{fig:typical-int}(b) gives the maximal drop height as a function of drop volume. 
Note, that there exists a minimal droplet volume $V_\mathrm{sn}$ given by the  saddle-node bifurcation in Fig.~\ref{fig:typical-int}(b). If one decreases the volume below $V_\mathrm{sn}$, the droplet collapses, i.e., it changes discontinuously into a flat film. The transition is hysteretic (first order) as the primary bifurcation at $V_\mathrm{c}$ is subcritical. The situation is different for freely evaporating droplets when the chemical potential is controlled instead of volume. For a more detailed comparison of the two cases see Ref.~\cite{Thie10}. 

\section{Parameter passing between particle-based model and continuum description}
\label{sec:pass}
%
The particle-based model is defined in terms of pairwise interactions
between beads, while the information that dictates the behavior of the
continuum description is the liquid-vapor tension, $\gamma$ and the
interface potential, $g(h)$. The latter quantifies the free-energy
cost of locating the liquid-vapor interface a distance $h$ away from
the solid substrate. Several strategies have been proposed to measure
the interface potential in computer simulation of particle-based
models: (i) The interaction between the interface and the substrate
can be obtained in the grandcanonical ensemble, where the chemical
potential $\mu$ controls the fluctuating thickness of the wetting
layer of the liquid on the substrate. The probability, $P(h)$, of
observing a wetting layer of thickness $h$ is related to the interface
potential via $g(h) = -k_{\rm B}T \ln P(h) +$ const
\cite{Muller2001,MM2003,LGM2006,JRE2010}, where the choice of
  the constant ensures the boundary condition
$g(h\to\infty)=0$. While being elegant, this computational technique
is limited to simple models because the grandcanonical ensemble
requires the insertion and deletion of polymers and concomitant
Monte-Carlo moves are only efficient for short polymers, low densities
or in the vicinity of the liquid-vapor critical point. (ii) A negative
curvature of the interface potential at a thickness $h$ signals the
spontaneous instability of a wetting layer. From the characteristic
length scale of this spinodal dewetting pattern one can deduce
information about $d^2 g(h)/dh^2$ \cite{Vrij1966,RS_SH_KJ_2001}. (iii) Here we use the pressure
tensor. This is a general technique that is not limited to short
polymers or low densities. It does not require the implementation of
particle insertion/deletion Monte-Carlo moves and can be
straightforwardly implemented in standard Molecular Dynamics program
packages.


%
\subsection{Virial pressure for a liquid film on a solid substrate}
\label{sec:pass:strat-slab}
%
\begin{figure}[t]
\includegraphics[width=0.8\hsize]{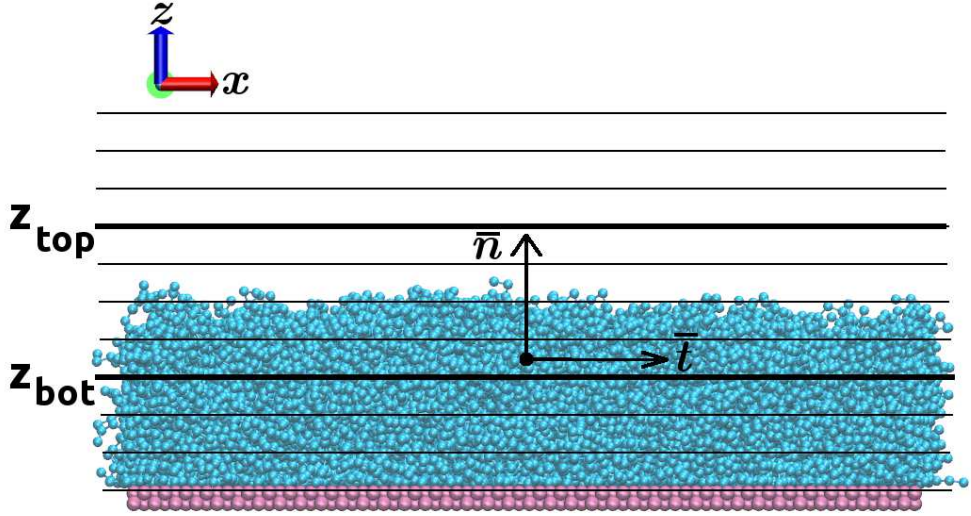}
\caption{Sketch of the slab geometry used to calculate the
  liquid-vapor interface tension $\gamma$. The pressure tensor components $p_\textrm{n}(z)$ and $p_\textrm{t}(z)$ are calculated in every slab $k$ and then their difference is integrated across the interface.}
\label{fig:p_tensor_slab}
\end{figure}

We study a supported polymer film as illustrated in Fig.~\ref{fig:p_tensor_slab} in the canonical ensemble. By virtue of the low vapor pressure of the polymer liquid, one can neglect evaporation effects. The flat liquid-vapor interface allows us to divide the system into thin parallel slabs (separated by the horizontal grey lines in Fig.~\ref{fig:p_tensor_slab}), whose normal vector $\vec{n}$ is perpendicular to the substrate. All relevant quantities can then be averaged over each slab, resulting in fields that depend on the $z$-coordinate only. 

In order to obtain the tension of the liquid-vapor and solid-liquid interfaces, $\gamma$ and $\gamma_{\rm SL}$, as well as the interface potential, $g(h)$, we consider a virtual change of the geometry of the simulation box such that the total volume $V$ remains unaltered. Using the scaling parameter $\lambda$, we relate the new linear dimensions, $L_{x}',L_{y}',L_{z}'$ of the simulation box to the original ones via $L'_x = \sqrt{\lambda}L_x,\; L'_y = \sqrt{\lambda}L_y,\; L'_z = \frac{1}{\lambda}L_z$. This scaling is the analog to the spreading of a droplet on a solid substrate. Thereby, only the liquid phase is subjected to this virtual change of the geometry but not the solid support.

The value $\lambda<1$ corresponds to a lateral squeezing of the liquid film on top of a solid substrate and a concomitant increase of the film thickness $h'=\frac{1}{\lambda}h$, where we have assumed that the liquid is incompressible. In the continuum model such a transformation gives rise to the following infinitesimal change of the canonical free energy \cite{deGennes04}
\begin{eqnarray}
\left.\frac{{\rm d}F(\lambda)}{{\rm d}\lambda}\right|_{\lambda=1} &=&
\left.\left[\gamma_{\rm SL}+\gamma+g(h)\right] 
\frac{{\rm d}L'_{x}L'_{y}}{{\rm d}\lambda}\right|_{\lambda=1}
+ \left.\frac{{\rm d}g(h)}{{\rm d}h} \frac{{\rm d}h'}{{\rm d}\lambda}\right|_{\lambda=1} L_{x}L_{y} \\
&=&  \left[\gamma_{\rm SL}+\gamma + g(h) - \frac{{\rm d}g(h)}{{\rm d}h}  h \right] L_{x}L_{y},
\label{eq:dfdl}
\end{eqnarray}
where, contrary to the related works in grandcanonical ensemble~\cite{BNR02,Han08}, we use the property of a canonical one and keep the number of particles in the liquid constant, i.e. constant volume $h L_x L_y = h' L'_x L'_y$ of the film  
\begin{equation}
 \frac{{\rm d}L'_x L'_y}{L'_x L'_y} + \frac{{\rm d}h'}{h'}=0.
\end{equation}

\begin{figure}[t]
\begin{minipage}[t]{0.44\textwidth}
\includegraphics[width=1.0\hsize]{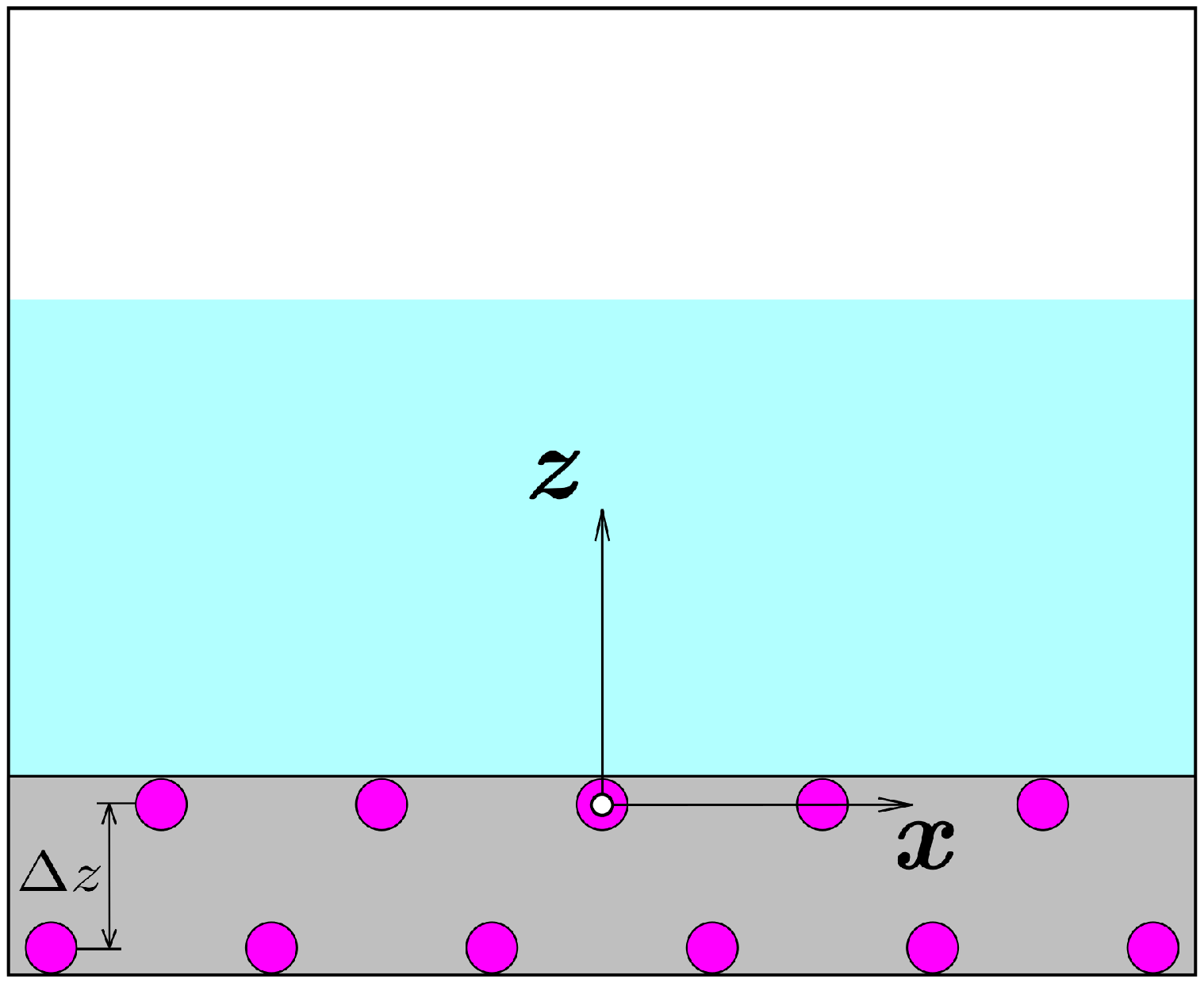}
\end{minipage}
\begin{minipage}[t]{0.1\textwidth}
\vspace{-2.7cm}\Huge{$\Rightarrow{}$}
\end{minipage}
\begin{minipage}[t]{0.44\textwidth}
\includegraphics[width=1.0\hsize]{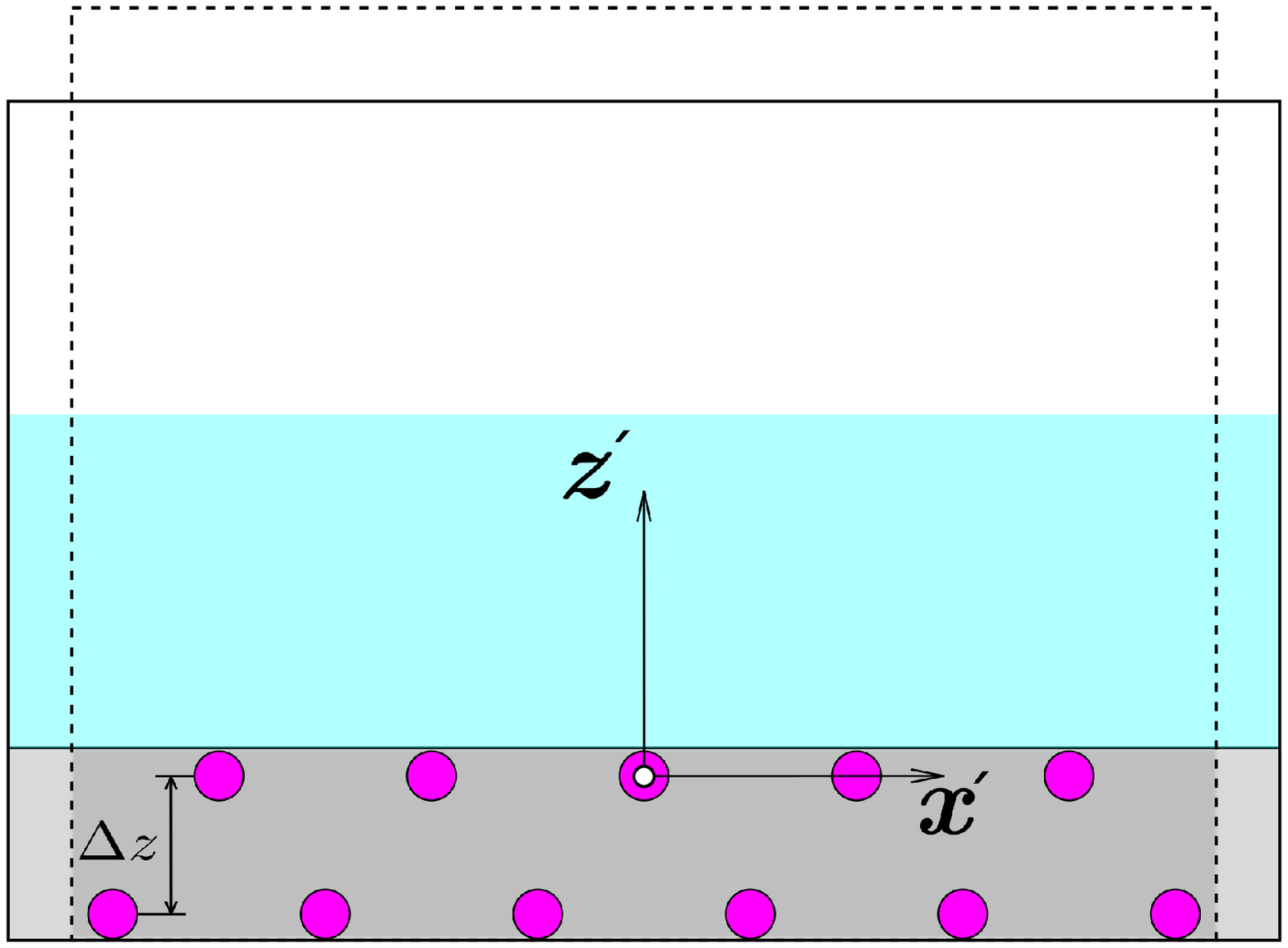}
\end{minipage}
\caption{A liquid (blue) in unscaled and scaled simulation boxes (left and right, respectively). Lila circles represent two layers of substrate atoms. The origin of $z$ axis is at the top layer of the substrate and for $x$ axis it is in the middle of the box. The scaling of the liquid phase is an analog to a spreading of the liquid on a supporting substrate. The entire substrate (shaded area) remains unscaled upon this virtual change of the geometry, preserving the distance $\Delta z$ between two atomic layers. $y$ axis is not shown for simplicity.} 
\label{fig:scaling}
\end{figure}
The scaling affects the beads of the polymeric liquid only, i.e., the lateral coordinates $x$ and $y$ are scaled by the factor $\sqrt{\lambda}$ and the normal component $z$ is scaled by $1/\lambda$. Upon scaling the liquid, the solid surface remains unaltered as indicated by shaded areas in Fig.~\ref{fig:scaling}. Therefore, the distance $\Delta z$ between two atomic layers and the coordinates of substrate particles are not changed. The origin of the coordinate system in $x$ and $y$ directions is taken in the middle of the simulation box, while in $z$ direction it is at the first layer of the substrate atoms.

In order to compute the change of free energy, we consider the canonical partition function
\begin{equation}
{\cal Z} = \frac{1}{n!\lambda_\textrm{T}^{3n}}\int \prod_{i=1}^n \; d^3\vec{r}_i \; 
   \exp\Big[- \beta\sum_{i<j}U(\vec{r_i}-\vec{r_j}) - \beta\sum_{s,i}U^\textrm{s}(\vec{r_i}-\vec{r_s})\Big]
\end{equation}
where $n$ is the number of particles in the system, $\beta =
\frac{1}{k_\textrm{B} T}$ and $\lambda_{T}$ is the thermal de-Broglie
wavelength. $U$ denotes the bonded and non-bonded interactions between
the polymer beads $i$ and $j$, and $U^\textrm{s}$ are the interactions between the polymer beads $i$ and the substrate particles $s$.

This separation of potentials allows us to express the partition function, ${\cal Z}(\lambda)$, of the scaled system through the scaling transformation of the original positions
\begin{eqnarray}
 {\cal Z}(\lambda) &= \frac{1}{n!\lambda_\textrm{T}^{3n}}\int \prod_{i=1}^n \; {\rm d}^3\vec{r}_i \;  \exp\Big[ &- \beta\sum_{i<j}U\left({\sqrt{\lambda}}(x_i-x_j),\,{\sqrt{\lambda}}(y_i-y_j),\,\frac1\lambda(z_i-z_j)\right) 
\\
&& - \beta\sum_{s_1,i}U^\textrm{s}\left({\sqrt{\lambda}}x_i-x_{s_{1}},\,{\sqrt{\lambda}}y_i-y_{s_{1}},\,\frac1\lambda z_i- z_{s_1}\right) 
 \nonumber \\
&& - \beta\sum_{s_2,i}U^\textrm{s}\left({\sqrt{\lambda}}x_i-x_{s_{2}},\,{\sqrt{\lambda}}y_i-y_{s_{2}},\,\frac1\lambda z_i- z_{s_2}\right)\Big],  \nonumber
\end{eqnarray}
where we explicitly separated the interaction of the polymer beads with the first and second layers of the unscaled substrate, $z'_{s_1} = z_{s_1}$ and $z'_{s_2} = z_{s_2} $. Differentiation with respect to $\lambda$ yields
\begin{eqnarray}
\frac{{\rm d}{\cal Z}}{{\rm d}\lambda} &= -\frac{\beta}{n!\lambda_\textrm{T}^{3n}}\int \prod_{i=1}^n \; {\rm d}^3\vec{r}_i \; \Big\{
  & \sum_{i<j} \Big(\frac{1}{2\lambda}(\frac{\partial U}{\partial x_{ij}}x_{ij} + \frac{\partial U}{\partial y_{ij}}y_{ij}) - \frac{1}{\lambda}\frac{\partial U}{\partial z_{ij}} z_{ij}\Big) 
  \\
&& +\sum_{s_{1},i} \Big(\frac{1}{2\lambda}(\frac{\partial U^\textrm{s}}{\partial x_{is_{1}}}x_{i} + \frac{\partial U^\textrm{s}}{\partial y_{is_{1}}}y_{i})  - \frac{1}{\lambda} \frac{\partial U^\textrm{s}}{\partial z_{is_{1}}} z_{i} \Big)  
 \nonumber \\
&& +\sum_{s_2,i}  \Big(\frac{1}{2\lambda}(\frac{\partial U^\textrm{s}}{\partial x_{is_{2}}}x_{i} + \frac{\partial U^\textrm{s}}{\partial y_{is_{2}}}y_{i})  - \frac{1}{\lambda} \frac{\partial U^\textrm{s}}{\partial z_{is_2}} z_{i}\Big)\Big\} \nonumber \\
&& \times \exp\Big[- \beta(\sum_{i<j}U + \sum_{s_{1},i}U^{\textrm{s}} + \sum_{s_{2},i}U^{\textrm{s}}) \Big] \nonumber
\end{eqnarray}
where $x_{ij}=x_{i}-x_{j}$. Then, in sums over correspondent substrate layers we replace the absolute coordinates of liquid particles by $z_{i}=z_{is_1} + z_{s_1}$ and $z_{i}=z_{is_2} + z_{s_2}$. Since the origin of the simulation box is chosen at the top layer of the substrate, we substitute $z_{s_1} = 0$ and $z_{s_2} = - \Delta z$. Therefore, we write the change of the free energy in the form
\begin{eqnarray}
\frac{{\rm d}F}{{\rm d}\lambda} \Big|_{\lambda = 1}&=& - k_{\rm B}T \frac{1}{\cal Z} \frac{{\rm d}{\cal Z}}{{\rm d}\lambda} \Big|_{\lambda = 1}\\
&=& \left \langle   \sum_{i<j} \Big(\frac{1}{2}(\frac{\partial U}{\partial x_{ij}}x_{ij} + \frac{\partial U}{\partial y_{ij}}y_{ij}) - \frac{\partial U}{\partial z_{ij}} z_{ij} \Big)      \right \rangle \nonumber\\
&& +  \left \langle \sum_{s,i} \Big(\frac{1}{2}(\frac{\partial U^\textrm{s}}{\partial x_{is}}x_{i} + \frac{\partial U^\textrm{s}}{\partial y_{is}}y_{i})  - \frac{\partial U^\textrm{s}}{\partial z_{is}} z_{is} \Big)   \right \rangle 
 +  \left \langle \sum_{s_2,i}  \frac{\partial U^\textrm{s}}{\partial z_{is_2}}  \right \rangle \Delta z  \\
 &=& \left \langle   \sum_{i<j} \Big( f_{z,ij} z_{ij} - \frac{1}{2}  \left(  f_{x,ij} x_{ij} +  f_{y,ij} y_{ij}\right) \Big)  \right \rangle \nonumber\\
&& +  \left \langle \sum_{s,i} \Big( f^\textrm{s}_{z,is}z_{is} - \frac{1}{2}( f^\textrm{s}_{x,is}x_{i} + f^\textrm{s}_{y,is}y_{i}) \Big) - \sum_{s_{2},i} f^\textrm{s}_{z,is_{2}}  \Delta z   \right \rangle
\label{eq:sl_total_corr}
\end{eqnarray}
where $f_{x,ij}$ denotes the $x$-component of the force acting between polymer beads, $i$ and $j$. $\langle \cdots\rangle$ denote averages in the canonical ensemble.

The first term of Eq.~(\ref{eq:sl_total_corr}) is the anisotropy of the pressure inside the liquid \cite{AllenTild89, FrenkSmit02}. Using the approach of Irving and Kirkwood~\cite{JI_JK_1950}, we define profiles of the normal and tangential pressure in a slab $k$ according to~\cite{PS_JH_1982, JW_DT_DR_JH_1983, JH_FS_1984, FV_JB_KB_2000}
\begin{equation} 
p_\textrm{n}(k)=k_\textrm{B}T\langle\rho(k)\rangle+\frac{1}{V_\textrm{sl}}\Big\langle{\sum_{i<j}}^{(k)} f_{z,ij} z_{ij} \, \eta_{k}(\vec{r}_{ij})\Big\rangle,
\label{eq:p_n_liq_liq}
\end{equation}
and
\begin{equation}	 
p_\textrm{t}(k)=k_\textrm{B}T\langle\rho(k)\rangle+\frac{1}{2V_\textrm{sl}}\Big\langle{\sum_{i<j}}^{(k)}  \left(  f_{x,ij} x_{ij} + f_{y,ij} y_{ij} \right) \, \eta_{k}(\vec{r}_{ij}) \Big\rangle,
\label{eq:p_t_liq_liq}
\end{equation}
where $\rho(k)$ is the number density in a slab $k$ and
$V_\textrm{sl}$ denotes the volume of the slab. The  sum
$\sum_{i<j}^{(k)}$ runs over particles $i$ and $j$ if the line
connecting them crosses the boundary of slab $k$ (then $
\eta_{k}(\vec{r}_{ij})$ is the fraction of that line that is located
in slab $k$) or if both particles are in slab $k$ (then $\eta_{k}(\vec{r}_{ij})=1$). 

Using this definition of the local pressure and
Eq.~(\ref{eq:dfdl}), we finally rewrite Eq.~(\ref{eq:sl_total_corr}) as
\begin{eqnarray}
\gamma_{\rm film}(h) \equiv \gamma_{\rm SL}+\gamma + g(h) - \frac{{\rm d}g(h)}{{\rm d}h}  h
&=& \int {\rm d}z\; \left[ p_{\rm n}(z) - p_{\rm t}(z)\right] \label{eq:res}\\
&& +  \frac{1}{L_{x}L_{y}}\left \langle \sum_{s,i} \Big[ f^\textrm{s}_{z,is}z_{is} - \frac{1}{2}( f^\textrm{s}_{x,is}x_{i} + f^\textrm{s}_{y,is}y_{i}) \Big] - \sum_{s_{2},i} f^\textrm{s}_{z,is_{2}}  \Delta z   \right \rangle \nonumber
\end{eqnarray}
The free energy per unit area of the supported polymer film is given by the anisotropy of the pressure in the liquid and contributions due to the direct interaction between the liquid and the solid substrate. In the limit that the substrate is laterally homogeneous the terms involving the lateral forces between solid and liquid vanish.

We particularly stress that in the canonical ensemble the difference of the film tension $\gamma_{\rm film}(h)$ and interface tensions $\gamma_{\rm SL}$ and $\gamma$ is not the interface potential $g(h)$~\cite{BNR02,HeHe10,Han08}, but of the form of Legendre transform $g(h) - h \frac{{\rm d}g(h)}{{\rm d}h}$.

%
\subsection{Solid-liquid and liquid-vapor interface tensions}
\label{sec:pass:sub-liq}
%

In the absence of a solid substrate, the liquid is separated by a
liquid-vapor interface from its coexisting vapor phase. In this
special case, Eq.~(\ref{eq:res}) simplifies and allows us to
  measure the liquid-vapor interface tension  through the anisotropy of the pressure tensor components across the interface as \cite{Tolman_1948,FV_JB_KB_2000, JW_DT_DR_JH_1983}:
\begin {equation}
  \gamma = \int_{z_\textrm{top}}^{z_\textrm{bot}} {\rm d}z\; [p_\textrm{n}(z) - p_\textrm{t}(z)]
  \label{eq:surf_ten_tolman}
\end {equation}
We find $\gamma = 0.512\pm0.006\epsilon/\sigma^2$ which agrees well
with previous calculations for similar systems
\cite{JS_MM_2008}. Mechanical stability requires that the normal
component of the pressure is constant throughout the system and equals
the coexistence pressure \cite{FV_JB_KB_2000}. Since the vapor
pressure of a polymer melt is vanishingly small, $p_{\rm n}(z) \approx
0$. We also note, that the anisotropy of the pressure is localized
around the interface and, therefore, the integration can be restricted
to an interval $[z_\textrm{bot},z_\textrm{top}]$ around the
interface. At the temperature of $k_{\rm B}T/\epsilon=1.2$ the coexistence density of the liquid inside a thick polymer film is $\rho_{0}\sigma^{3}=0.786$.

\begin{figure}[t]
 \includegraphics[width=0.8\hsize]{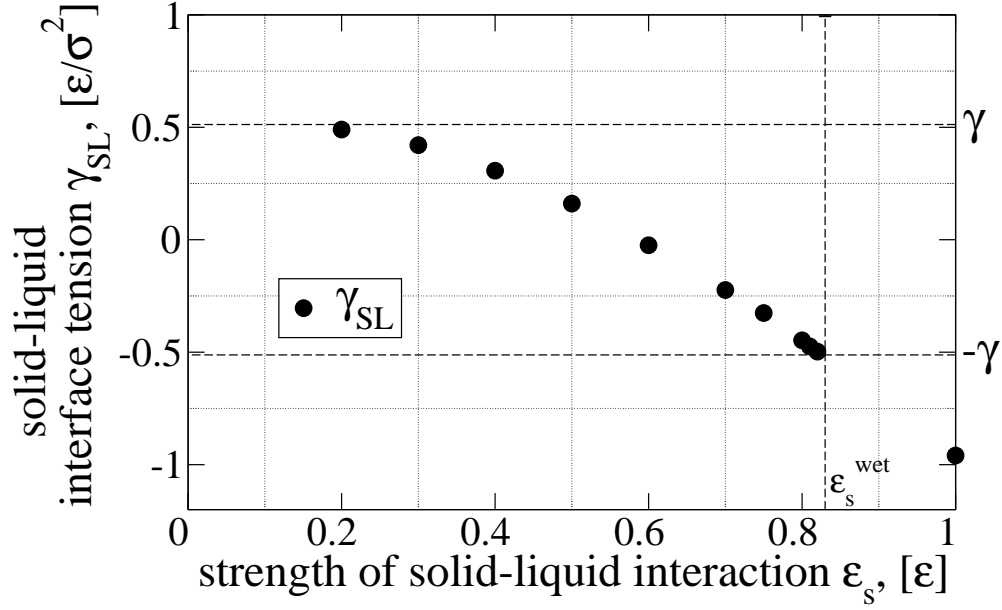}
 \caption{The dependence of solid-liquid interface tension
   $\gamma_\textrm{SL}$ on the strength of solid-liquid
   interaction $\epsilon_\textrm{s}$. The horizontal dashed lines
   represent the value of the liquid-vapor interface tension ($\gamma$) and the value of solid-liquid interface tension corresponding to the wetting transition (-$\gamma$). The wetting transition is localized at $\epsilon^\textrm{wet}_\textrm{s} \approx 0.83 \epsilon $.}
 \label{fig:SL_surf_tens}
\end{figure}

If we consider a liquid film in contact with the solid substrate, we can measure the solid-liquid interface tension $\gamma_{\rm SL}$ according to Eq.~(\ref{eq:res}) (provided that the thickness of the liquid film is sufficiently large to prevent the interaction of liquid-vapor and solid-liquid interfaces, i.e., $|g| \ll \gamma_{\rm SL}$). Like in the case of the liquid-vapor interface, the anisotropy of the pressure, as well as the additional contribution due to the interaction between the liquid and the solid, are localized in a narrow region near the interface between the polymer liquid and the solid. The solid-liquid interface tension depends on the strength $\epsilon_{\rm s}$ of the attractive interaction between solid and polymer liquid. The simulation results are presented in Fig.~\ref{fig:SL_surf_tens}. 

If the droplet on a substrate depicted in Fig.~\ref{fig:typical_int} is at equilibrium, one may describe the equilibrium of forces acting on its contact line by the macroscopic Young-Laplace equation that relates the interface energies and the equilibrium contact angle $\theta_\textrm{E}$ \cite{Young_1805,Lapl1806},
\begin {equation}
  \gamma_\textrm{SL} + \gamma \cos\theta_\textrm{E}- \gamma_\textrm{SV} = 0.
  \label{eq:young}
\end {equation}

Since the vapor pressure is vanishingly small for our polymer melt, we can neglect the interface tension between the solid substrate and the vapor phase, $\gamma_{\rm SV} \approx 0$ to a first approximation. Using this approximation, we find that the wetting and drying transitions occur at $\gamma_{\rm SL}(\epsilon_{\rm s}) \approx - \gamma$ and $\gamma_{\rm SL}(\epsilon_{\rm s}) \approx \gamma$, respectively. From the data in Fig.~\ref{fig:SL_surf_tens} we locate the wetting transition at $\epsilon^\textrm{wet}_\textrm{s} \approx 0.83 \epsilon$ and the contact angle reaches  $180^{o}$ for small values of $\epsilon_{\rm s}<0.2\epsilon$.

%
\subsection{Solid-vapor interface tension}\label{sec:pass:sub-vap}
%
\begin{figure}[t]
 \includegraphics[width=0.8\textwidth]{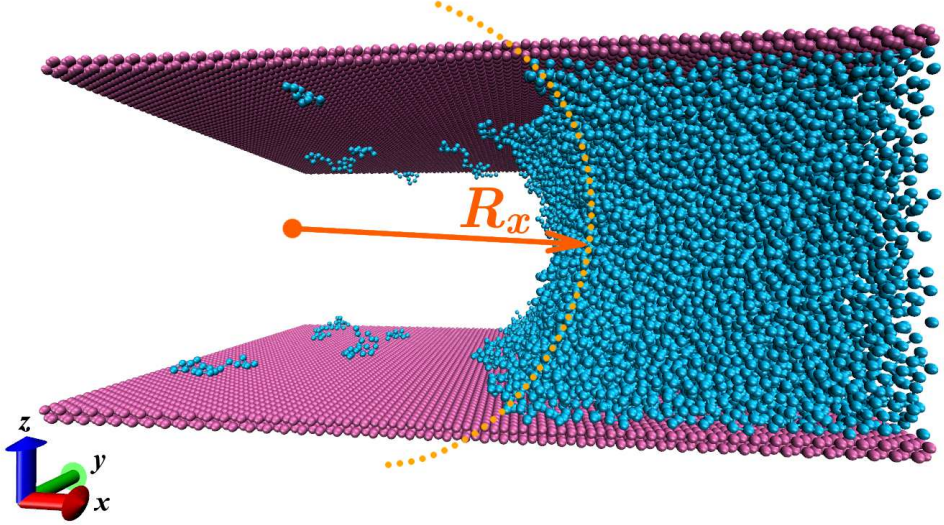}
 \caption{A part of a system used to determine solid-vapor interface
   tension $\gamma_\textrm{SV}$. The droplet serves as a reservoir to
   the chains adsorbed on the substrate. The yellow dotted line
   indicates the curvature of the liquid-vapor interface. The radius
   of curvature $R_x$ is indicated by the orange arrow.}
 \label{fig:meniscus}
\end{figure}

\begin{table}
 \begin{tabular}{l|c|c|c|c}
    & $\epsilon_\textrm{s} = 0.75\epsilon$ & $\epsilon_\textrm{s} = 0.80\epsilon$ & $\epsilon_\textrm{s} = 0.81\epsilon$ & $\epsilon_\textrm{s} = 0.82\epsilon$ \\
    \hline
    $\gamma_\textrm{SV}, [\epsilon / \sigma^2]$ & 0 & -0.00281 & -0.00475 & -0.00523 (-0.01642) \\
    \hline
    $\gamma_\textrm{SL}, [\epsilon / \sigma^2]$ & -0.32576 & -0.44737 & -0.47419 & -0.49761 \\
    \hline
    $\theta_\mathrm{E_{0}}$ (at $\gamma_\textrm{SV} = 0$), [degree]& 50.50 & 29.14 & 22.20 & 13.69 \\
    \hline
    $\theta_\textrm{E}$, [degree]& 50.50 & 29.77 & 23.57 & 15.98 (20.03) \\
 \end{tabular}
 \caption{Interface tensions of solid-vapor and solid-liquid
   interfaces and contact angles with ($\theta_\textrm{E}$) and
   without ($\theta_\mathrm{E_{0}}$) taking the solid-vapor interface tension
   into account. For $\epsilon_\textrm{s} =0.82 \epsilon$ the value
   $\gamma_{\rm SV}$ is affected by the finite value of $\Delta p$ and
   we  provide in parentheses an alternative estimate of the contact angle.}
 \label{tab:gamma_SV}
\end{table}

While the approximation $\gamma_{\rm SV} \approx 0$ is appropriate for small values of the strength of attractive solid-liquid interactions, $\epsilon_{\rm s}$, the quality of this approximation deteriorates in the vicinity of the wetting transition. If the wetting transition were of second-order, the amount of liquid adsorbed onto the substrate, would continuously diverge as we approach the wetting transition. Even for a first-order wetting transition we expect that the adsorbed amount (i.e., the film thickness $h_{\rm min}$ at which the interface potential exhibits a minimum) will increase when $\epsilon_{\rm s}$ increases towards its transition value. In this case the approximation $\gamma_{\rm SV} \approx 0$ becomes unreliable and we employ  a meniscus geometry as shown in Fig.~\ref{fig:meniscus} to extract the value of the solid-vapor tension. 

The film thickness is chosen sufficiently large, such that the deviation of the pressure from its coexistence value, $\Delta p \sim -\frac{\gamma}{R_x+R_y}$, with $R_{x}$ and $R_{y}=\infty$ denoting the principle radii of curvature of the meniscus, has only a small influence on the adsorbed amount of polymer and $\gamma_{\rm SV}$. 
Since $\Delta p<0$, the adsorbed amount in the simulations will be smaller than at coexistence, $\gamma_{\rm SV}$ will be too large (i.e., negative $\gamma_{\rm SV}$ will have an absolute value that is too small), and we will slightly underestimate the contact angle, $\theta_{\rm E}$. This correction to the deviation of the approximation $\gamma_{\rm SV} \approx 0$, however, is insignificant for the used system size for all values of $\epsilon_{\rm s}$ but the close vicinity of the wetting transition $\epsilon^\textrm{wet}_\textrm{s} \approx 0.83 \epsilon$. Therefore, at $\epsilon_\textrm{s} = 0.82 \epsilon$, we have used an alternative method as described in the following Sec.~\ref{sec:pass:pot}.

For the calculation of $\gamma_\textrm{SV}$ we used the same procedure as earlier for the solid-liquid interface tensions of a film, but the procedure is only applied to the part of the simulation box that is far away from the meniscus-forming liquid bridge. The values of $\gamma_\textrm{SV}$ and $\gamma_\textrm{SL}$ (for comparison) are presented in Table~(\ref{tab:gamma_SV}). One notices the increase in $\gamma_\textrm{SV}$ when the wetting transition is approached. However, compared to the influence on the solid-liquid interface tension the effect is small.  Nevertheless, it becomes more important the closer one comes to the wetting transition, and the correction of the contact angles is significant when one compares profiles of drops of different sizes with the prediction of Eq.~(\ref{eq:young}).

We compare the shape of drops obtained from the particle-based and continuum description in the vicinity of the wetting transition. In the following detailed comparison we employ the values
$\epsilon_\textrm{s} =0.75 \epsilon$, $\epsilon_\textrm{s} =0.80 \epsilon$, and $\epsilon_\textrm{s} =0.81 \epsilon$ for the solid-liquid interaction strength that correspond to contact angles $\theta_\mathrm{E}=50.50^\mathrm{o}$, $\theta_\mathrm{E}=29.77^\mathrm{o}$, $\theta_\mathrm{E}=23.57^\mathrm{o}$, respectively. For  $\epsilon_\textrm{s} =0.82 \epsilon$, however, we will use the more accurate value,  $\theta_\mathrm{E}=20.03^\mathrm{o}$, extrapolated from the interface potential instead.

%
\subsection{Interface potential and Derjaguin pressure}
\label{sec:pass:pot}
%

\begin{figure}[t]
\includegraphics[width=0.8\hsize]{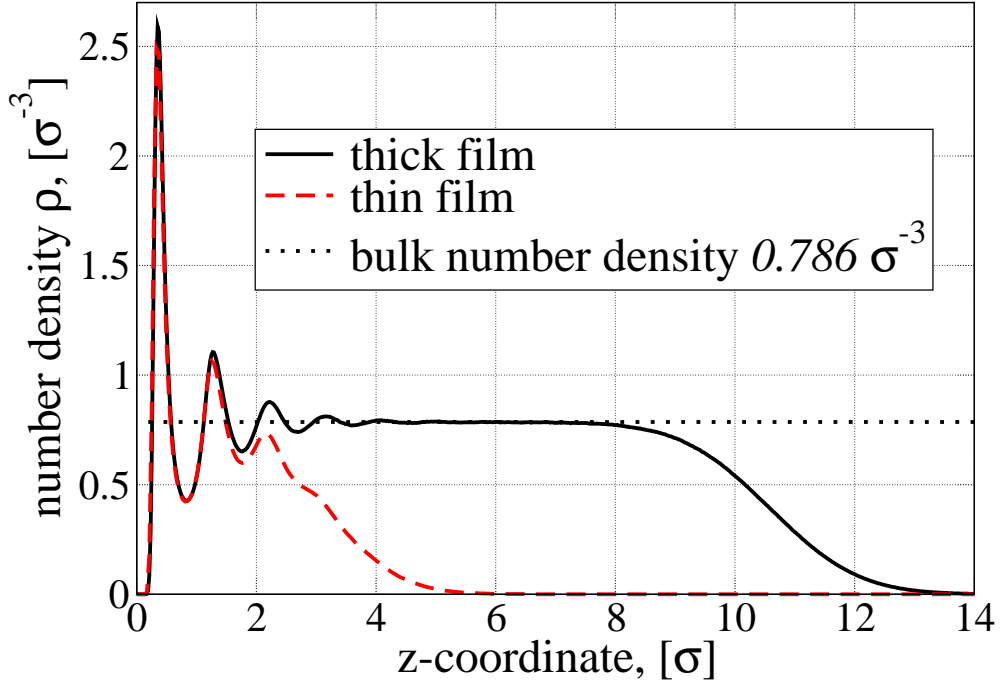}
\caption{Density profiles of a polymer film at $\epsilon_\textrm{s} = 0.80\epsilon$. The solid black line represents a thick film with a bulk region separating solid-liquid and liquid-vapor interfaces. In the case of a thin film (dashed red line) it is no longer possible to distinguish the two interfaces. The dotted horizontal line indicates the coexistence number density.}
\label{fig:dens_thick_and_thin_films}
\end{figure}

If we consider a polymer film on top of the solid substrate,
Eq.~(\ref{eq:res}) provides information about the solid-liquid and
liquid-vapor interface tensions, $\gamma_{\rm SL}$ and $\gamma$ as
well as the interface potential, $g(h)$. For a thick film
(cf.~Fig.~\ref{fig:dens_thick_and_thin_films}), the transitions
  in polymer density at the two interfaces are well separated, and the
  density at the center of the film approaches the bulk coexistence
  value. In this case, also the contributions to Eq.~(\ref{eq:res})
  that stem from the two interfaces can be well separated. The
  anisotropy of the pressure tensor at the solid substrate gives
  $\gamma_{\rm SL}$, and the one at the liquid-vapor interface gives
  $\gamma$. Thus, the interface potential vanishes, $g(h \to \infty)
  \to 0$, indicating that the liquid-vapor interface will not interact
  with the substrate if the film is sufficiently thick.

However, upon decreasing the film thickness, the two interfaces start to interact and
the contributions of the solid-liquid and liquid-vapor interfaces can
not be separated anymore. The interaction between the interfaces is
quantified by the interface potential, $g(h)$, or equivalently, by the
Derjaguin pressure $\Pi(h) = -\frac{{\rm d} g(h)}{{\rm d} h}$. From
Fig.~\ref{fig:dens_thick_and_thin_films} we observe that for small
film thickness both interface density profiles are distorted, and the density does not reach its coexistence value at the center of the film. The distortion of the density profile far away from the interfaces is characterized by the bulk correlation length, $\xi_{0}$, which therefore sets the length scale of the interface potential \cite{Schick90}. 

Since we have determined $\gamma_{\rm SL}$ and $\gamma$ independently,
we are able to extract the interface potential, $g(h)$, from the
simulation data for thin films. To this end, we have to define
  the location of the liquid-vapor interface, i.e., the film
  thickness, $h$. There are several options: Either (i) one
determines the position where the density equals a predefined value,
typically $(\rho_\mathrm{liq}+\rho_\mathrm{vap})/2$ (crossing
criterion) or (ii) one defines the film thickness via the adsorbed
excess (Gibbs dividing surface),

\begin{equation}
\Delta \Gamma_\mathrm{ads} = L_{x}L_{y}\int \mathrm{d}z\;
\left[\rho(z)-\rho_\mathrm{vap}\right] \equiv
\left[\rho_\mathrm{liq}-\rho_\mathrm{vap}\right]  L_{x}L_{y}h
\label{eq:Gads}
\end{equation}  

In this work we adopt the integral criterion (\ref{eq:Gads}) to define the film thickness. Neglecting the vanishingly small vapor density at coexistence, we obtain

\begin{equation}
h_\mathrm{eff} = \frac{N_\textrm{mon}}{\rho_\mathrm{liq} A_\textrm{film}}, 
\end{equation}
where $N_\textrm{mon}$ is the number of monomers of the liquid inside the simulation box and $A_\textrm{film}=L_{x}L_{y}$ is the area of the substrate underneath the film.

We note that both definitions become problematic for film thicknesses where the curvature of the interface potential is negative, $\frac{{\rm d}^{2}g}{{\rm d}h^{2}}<0$. In this regime of film thicknesses a laterally extended, homogeneous film becomes unstable with respect to spinodal dewetting \cite{Vrij1966,Mitl93,Thie10}).  However, even in this film thickness region, the films can be linearly or even absolutely stable if the lateral extension of the simulation box is sufficiently small. The related critical values depend on film thickness \cite[see, e.g., Fig.8 of]{TVNP01}. In the simulation, we can still obtain meaningful data for the interface potential if we restrict the lateral system size to be smaller than the characteristic wavelength of the spontaneous rupture process.

Additionally, we mention that the liquid-vapor interface in our
Molecular Dynamics simulations exhibits local fluctuation of its
height (i.e., capillary waves), and the Gibbs dividing surface
measures the laterally averaged film thickness. The interaction of the
liquid-vapor interface with the substrate imparts a lateral
correlation length, $\xi_{\|} = 2\pi \sqrt{\gamma/\frac{{\rm
      d}^{2}g}{{\rm d}h^{2}}}$, onto these interface
fluctuations. These fluctuations give rise to a weak dependence of the
interface potential on the lateral system size for
$L_{x},L_{y}<\xi_{\|}$, i.e., the interface potential is renormalized
by interface fluctuations. Qualitatively, the effect of fluctuations
is to extend the range of the potential, i.e., $\xi =
\xi_{0}(1+\omega/2)$ with $\omega=\frac{k_{\rm B}T}{4\pi
  \xi_{0}^{2}\gamma}$ \cite{Lipowsky87}.

The interface potential exhibits a minimum at small film thickness, $h_{\min}$. This film thickness characterizes the amount of liquid adsorbed on the substrate in contact with the vapor. As illustrated in Sec.~\ref{sec:pass:sub-vap} $\gamma_\textrm{SV}=0$ and therefore no chains are adsorbed on the substrate except for the close vicinity of the wetting transition. The free energy of such a vanishingly thin polymer film is given by $\gamma_\textrm{film}(h_\textrm{min})=\gamma_{\rm SL} + \gamma + g(h_{\min})=\gamma_{\rm SV}$. Thus, the measurement of the different tensions for a planar polymer film provides the value of $g(h_{\rm min})$.  

Alternatively, we can use the measured value $g(h_{\min})$, in turn,
to estimate the solid-vapor tension, $\gamma_{\rm SV}$. We have
employed this strategy for $\epsilon_\textrm{s} = 0.82 \epsilon$,
where the finite curvature of the meniscus result in a relevant deviation of the pressure from its coexistence value. Extrapolating the simulation data to the thickness $h_{\min} \approx 0$ we obtain $\gamma_{\rm SV}= -0.01642$. We will use this more accurate value, which is not affected by the curvature of the meniscus and that is compatible with the interface potential, in the comparison with the continuum model in Sec.~\ref{sec:res}.

Since Eq.~(\ref{eq:res}) only provides the Legendre
  transformation of the interface potential and we require an
analytical expression for the continuum model, we make an Ansatz for
the functional form of $g(h)$. Generally, one can distinguish between
short-range and long-range contributions to the interface potential \cite{dietr88, Schick90}. The long-range contribution results from dispersion forces
between the liquid and the substrate. In our particle-based model, however, we do only consider the short-range part as our LJ interaction~(\ref{eq:lj}) is cut off at $r_c$.  Thus, there is no long-range contribution in our model in contrast to previous works, when an effective long-range contribution was taken into account despite finite interaction cut off~\cite{BNR02,Han08}. The short-range contribution to $g(h)$ stems from the distortion of the interface profile due to the nearby presence of the solid substrate as illustrated in Fig.~\ref{fig:dens_thick_and_thin_films}, and it is typically expanded in a series of exponentials \cite{dietr88,Schick90}
\begin{equation}
g_\textrm{sr}(h) = a e^{-h/\xi} - b e^{-2h/\xi} + c e^{-3h/\xi} - d e^{-4h/\xi} + \dots
\label{eq:sr}
\end{equation}

\begin{figure}[t]
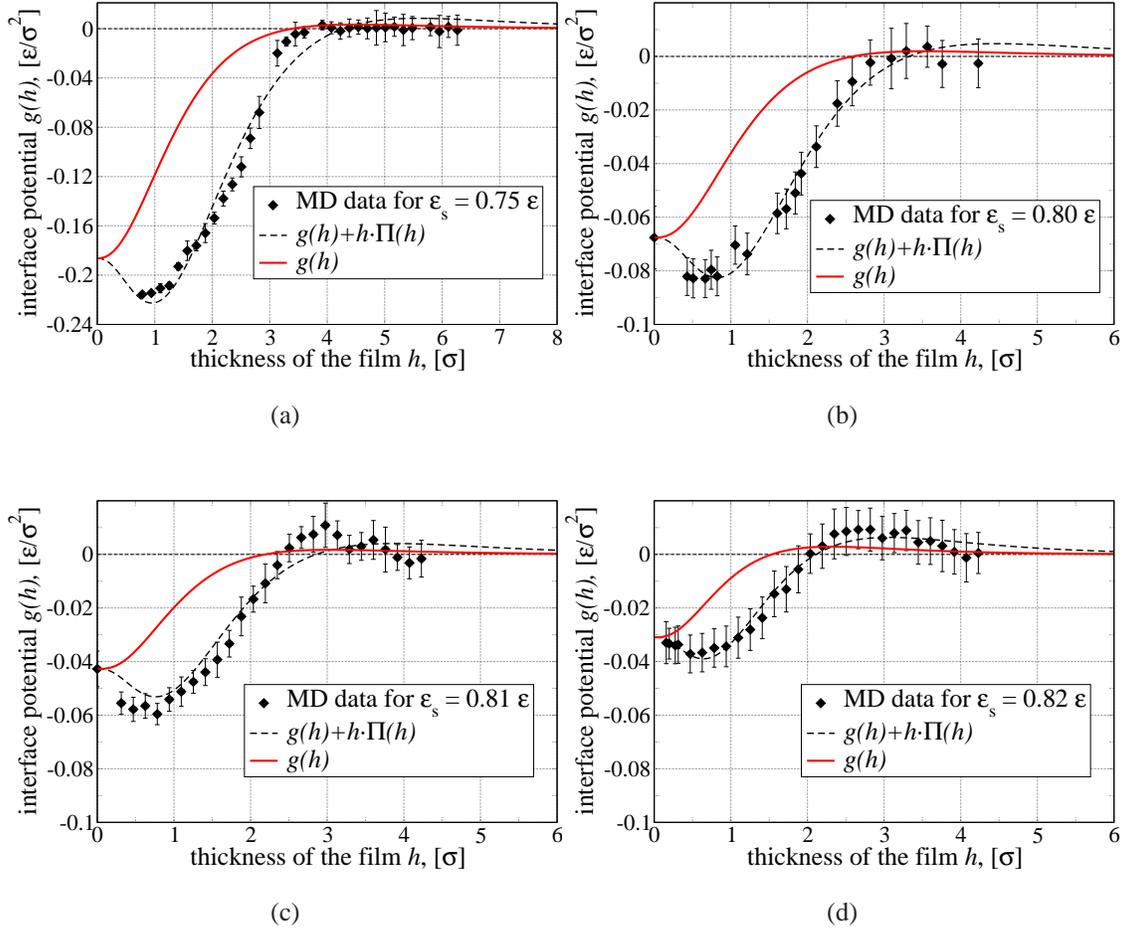

\subfloat[]{\label{fig:fit_075}\includegraphics[width=0.45\hsize]{Fig08a_int_pot_eps_075_0.eps}}
\subfloat[]{\label{fig:fit_080}\includegraphics[width=0.45\hsize]{Fig08b_int_pot_eps_080_0.eps}}\vspace*{0.4cm}

\subfloat[]{\label{fig:fit_081}\includegraphics[width=0.45\hsize]{Fig08c_int_pot_eps_081_0.eps}}
\subfloat[]{\label{fig:fit_082}\includegraphics[width=0.45\hsize]{Fig08d_int_pot_eps_082_1.eps}}
\caption{Panels (a), (b), (c) and (d) give the interface potential $g(h)$ at $\epsilon_\textrm{s} = 0.75 \epsilon$, $\epsilon_\textrm{s} = 0.80 \epsilon$, $\epsilon_\textrm{s} = 0.81 \epsilon$ and $\epsilon_\textrm{s} = 0.82 \epsilon$,
  respectively. They are obtained by fitting the MD results for the
  tension $\gamma_{\rm film} (h) - \gamma - \gamma_\textrm{SL}$ of films of various small thicknesses
  (black symbols with error bars) by the expression $g(h) + h \Pi(h)$ (dashed black line)
  obtained employing the first four terms of the short-range part of
  the interface potential $g_\textrm{sr}(h)$. The resulting interface
  potential $g(h)$ is given as solid red line.
  Note, that the minimal value of $g_\mathrm{min}$
  is always reached at vanishingly small thicknesses $h \approx 0
  \sigma$, as there is no precursor film in our MD model.}
\label{fig:int_pot}
\end{figure}

In order to obtain $g(h)$ in practice, we fit its Legendre transform $g(h)-h \frac{{\rm d}g}{{\rm d}h}$ by a sum of four exponential terms like in Eq.~(\ref{eq:sr}), and enforce that the interface potential exhibits a minimum at $h_{\rm min} \approx 0$ (there is no precursor film in our MD model) with a value $g(h_{\min})$, as obtained by the measurement of the interface tensions. The resulting fits for $g(h)$ at $\epsilon_\textrm{s} = 0.75 \epsilon, 0.80 \epsilon, 0.81 \epsilon$ and $0.82 \epsilon$ are given as solid lines in Figs.~\ref{fig:fit_075}-\ref{fig:fit_082}. The parameters of the fits are presented in Table~(\ref{tab:fits}).

\begin{table}
 \begin{tabular}{l|c|c|c|c}
    Parameter & $\epsilon_\textrm{s} = 0.75\epsilon$ & $\epsilon_\textrm{s} = 0.80\epsilon$ & $\epsilon_\textrm{s} = 0.81\epsilon$ & $\epsilon_\textrm{s} = 0.82\epsilon$ \\
    \hline
    $a, [\epsilon / \sigma^2]$ & 0.13191 & 0.06485 & 0.05057 & 0.06132 \\
    \hline
    $b, [\epsilon / \sigma^2]$ & 1.40871 & 0.58700 & 0.41256 & 0.36875 \\
    \hline
    $c, [\epsilon / \sigma^2]$ & 1.67606 & 0.70902 & 0.50249 & 0.42963 \\
    \hline
    $d, [\epsilon / \sigma^2]$ & 0.58566 & 0.25447 & 0.18323 & 0.15318 \\
    \hline
    $\xi, [\sigma]$ & 1.51770 & 1.26964 & 1.14735 & 1.00512 \\
 \end{tabular}
 \caption{Parameters of the fitting curves of $g(h)$ for the case where the
   first four terms of the short-range contributions
   [Eq.~(\ref{eq:sr})] are taken into account. Note, that only three
   parameters are independent since there are two additional
   constraints: The local minimum criterion at $h \approx 0 \sigma$
   implies $d = (a-2b+c)/4$ and the Young-Dupr\'{e} relation \ref{eq:young-dupre}
   dictates the value $g_\mathrm{min}$ by setting 
$b = a + c - d - g_\mathrm{min}= 1.5a + 0.5c - 2g_\mathrm{min}$.}
 \label{tab:fits}
\end{table}
Using  the macroscopic Young-Dupr\'{e} relation, one observes that value of the minimum of $g(h)$ dictates the contact angle \cite{deGennes04}
\begin{equation}
 g(h_{\mathrm{min}}) = \gamma(\cos\theta_\textrm{E} - 1) 
 \label{eq:young-dupre}
\end{equation}
Much more information can be extracted from the interface potential:
(i) The shape of the interface potential  controls deviations of the
drop shape from a spherical cap in the vicinity of the wetting
transition. (ii) Within the square-gradient approximation the integral
of $\sqrt{g(h)}$ is related to the line tension at the three-phase
contact line \cite{Indekeu1992,MN_SD_1993,TG_SD_1998,LS_MN_SD_2007}. For all values of $\epsilon_{\rm s}$
investigated in the particle-based model, the line tension is expected
to be negative. (iii) The observation that $g(h)$ increases above zero at intermediate values of $h$ indicates that the wetting transition is of first-order.

\section{Static case - Sitting droplets}
\label{sec:res}
%

\begin{figure}[t]
 \includegraphics[width=1.0\textwidth]{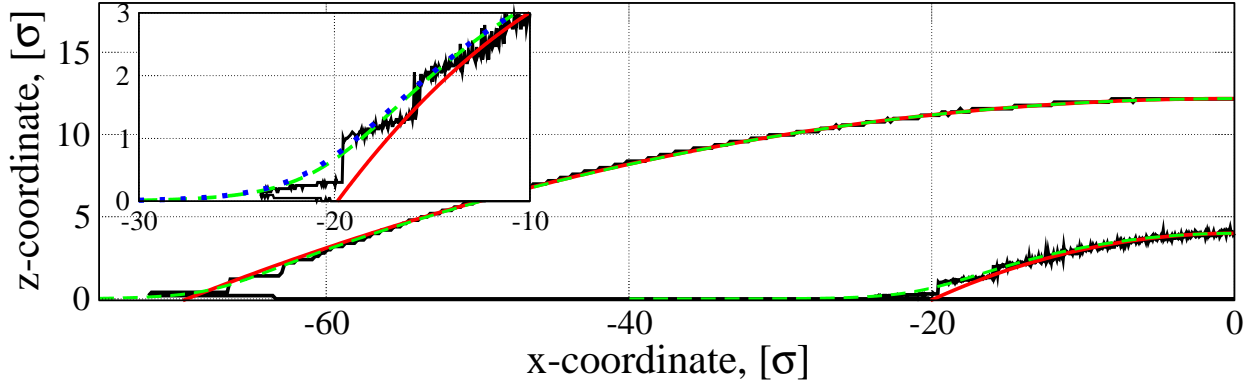}
 \caption{(color online) Profiles of two-dimensional droplets obtained
   by cutting cylindrical droplets obtained in MD simulations [solid
   noisy line (black online)] for the case
   $\epsilon_\mathrm{s}=0.82\epsilon$ for two values of $h_{max}$
   ($4.046\sigma$ and $12.181\sigma$). The
   corresponding spherical cap fit is given as solid smooth line (red
   online).  The MD drops are compared with results of the continuum
   model Eq.~(\ref{eq:el}) with the full curvature
   [Eq.~(\ref{eq:pi-full})] and in long-wave approximation
   [Eq.~(\ref{eq:pi-longwave})] that are given as dashed (green
   online) and dotted (blue online) lines, respectively.  The inset
   shows a zoom into the three-phase contact line region of the
   smaller droplet.}
  \label{fig:drop_prof_comp082}
\end{figure}

\begin{figure}[t]
 \includegraphics[width=0.95\textwidth]{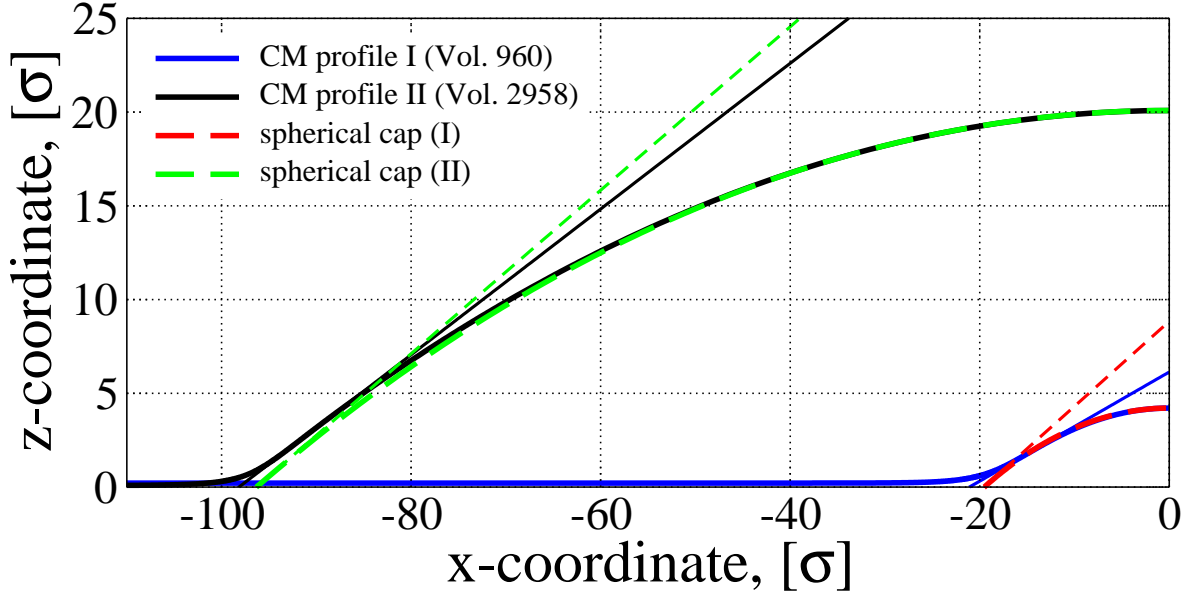}
  \caption{(color online) Droplet profiles as obtained from continuum theory
            with full curvature (heavy solid lines) for $\epsilon_\mathrm{s} =
            0.81 \epsilon$ and drop height (a) $H=4$ and (b) $H=20$. Also shown
            are the spherical caps as obtained from the curvature at
            the drop maxima (heavy dashed lines), and the tangent
            lines at the point of the steepest slope of the profile
            (thin solid line), and the tangent line of the spherical
            cap profile at precursor height (thin dashed line). Drop
            height $H$ is defined as difference of height at maximum
            and precursor height.}
  \label{fig:drop-prof-thinfilm}
\end{figure}

In the following we will compare the shape of droplets obtained from
the particle-based model and the continuum description. This
comparison focuses on droplets with small contact angles
$\le50^\mathrm{o}$ obtained in the particle-based model for strengths
of solid-liquid interaction close to the wetting transition
($\epsilon_\textrm{s} = 0.75 \epsilon$  to $0.82 \epsilon$). Different
numbers of polymer chains are used to create cylindrical 2d droplets
(3d ridges) of varying volumes and hence heights. Data are sampled
with a frequency of 4000 MD steps. This time interval  between two
samples corresponds to the Rouse relaxation time for a similar polymer
liquid $\tau_R = 25.6\,\pm\,5\,\tau$  \cite{JS_MM_2008}. For small
droplets (up to 600 chains) the sampling lasted $2\times10^6$ steps,
whereas for bigger ones (up to 9600 chains) this interval was
increased up to $10^7$ steps, because large fluctuations of the
droplet shape occur. As a result, every density profile is obtained by averaging over $500$ (small 
drops) to $2500$ (large drops) snapshots. To extract the droplet shape and measure the contact angle, we use a set of density profiles obtained in 10 independent runs. In total, all large droplets are simulated over $10^8$ steps.

The resulting cylindrical droplet snapshots are cut into slices along the invariant $y$-direction. In every slice the two-dimensional ($x,z$) density map is created with respect to the center-of-mass of the droplet cut. An average over these maps results in the average number density profile in the $(x,z)$ plane. A two-dimensional drop profile is extracted by localizing the solid-liquid and liquid-vapor interfaces by the crossing criterion for the density as $\rho_\mathrm{int}=(\rho_\mathrm{liq}+\rho_\mathrm{vap})/2$.  Examples of profiles are presented in Fig.~\ref{fig:drop_prof_comp082}. The resulting profiles are then compared to the ones extracted from the employed continuum models, which are also presented in Fig.~\ref{fig:drop_prof_comp082}.

One popular characteristics of the drop shape is the contact angle,
because it is related to the balance of interface tensions at the
three-phase contact line of a macroscopic drop. For finite-sized
drops, however, the contact angle is not uniquely defined: (i) One may
define a \textit{mesoscopic contact angle} $\theta_{\mathrm {mes}}$ as
the slope at the inflection point of the droplet profile. This is
often done in thin film models \cite{Thie01,DT_UT_LP_2012}, however,
the steepest slope obtained in this way may not coincide with the
(larger) macroscopic contact angle even in the limit of large drop
size \cite{Shar93b}. This corresponds to the distinction of
  macroscopic and microscopic contact angle in Ref.~\cite{Star92}.
Moreover, in the particle-based model, the inflection point may be located very close to the three-phase contact line where liquid-like layering effects of the particle fluid may occur and affect the drop profile \footnote{Similar 			extrapolation schemes, like defining a contact angle via
	the steepest slope of the liquid-vapor interface or the local extrapolation
	of the droplet shape towards the contact line, have been explored
	for the particle-based model but did not give reliable results for
	the contact angle.}.
(ii) Alternatively, one may define a \textit{spherical cap contact angle} by approximating the drop profile by a spherical cap profile with a minimal radius of 
curvature $R = -1/\kappa$, i.e., using the curvature at
$h_\mathrm{max}$. The resulting contact angle is $\theta_{\mathrm
  {sph}}=\arccos \left(1-h_{\mathrm {max}}/R\right)$
\footnote{There are different strategies of measuring the contact angle with
	the spherical cap approximation: direct geometric measurements~\cite{LMcD_MM_KB_2002,GB_ER_2008},
	estimation from the center of mass position~\cite{JS_MM_2008} or
	from the volume of the droplet. In our MD simulations, we define
	a contact angle by the geometrical method. Other methods give a similar
	result as all of them assume the spherical shape of the droplet.}.
In the profiles extracted from the particle-based
model, we extract $\theta_{\mathrm {sph}}$ by only considering the
central part of the drop to define the curvature. In this way, the
calculation is not perturbed by liquid-like layering effects or by the
short-range interface potential that distorts the liquid-gas
interface close to the three-phase contact line. The height of the drop is determined as the difference of the highest point of the spherical cap and the position of the solid-liquid interface. $\theta_{\mathrm {sph}}$ converges to the proper macroscopic contact angle in the limit of large drop size, but may misrepresent the shape and volume of small droplets. 

For the continuum model, the two angles $\theta_{\mathrm {sph}}$ and
$\theta_{\mathrm {mes}}$ are illustrated in
Fig.~\ref{fig:drop-prof-thinfilm} that shows two droplet profiles
$h(x)$ as obtained from Eqs.~(\ref{eq:el}) with (\ref{eq:pi-full}),
their approximated spherical cap profiles and the tangents of
  $h(x)$ at the point of steepest slope (giving $\theta_{\mathrm
    {mes}}$) and of the spherical cap profile at the point where it
  crosses the precursor height (giving $\theta_{\mathrm {sph}}$).
 One clearly notes that the two measures differ, and that the difference decreases with increasing droplet size. We will see below that the two measures do not converge even for very large drops. In the following we focus on the spherical cap contact angle $\theta_{\mathrm {sph}}$. 
 
\begin{figure}[t!]
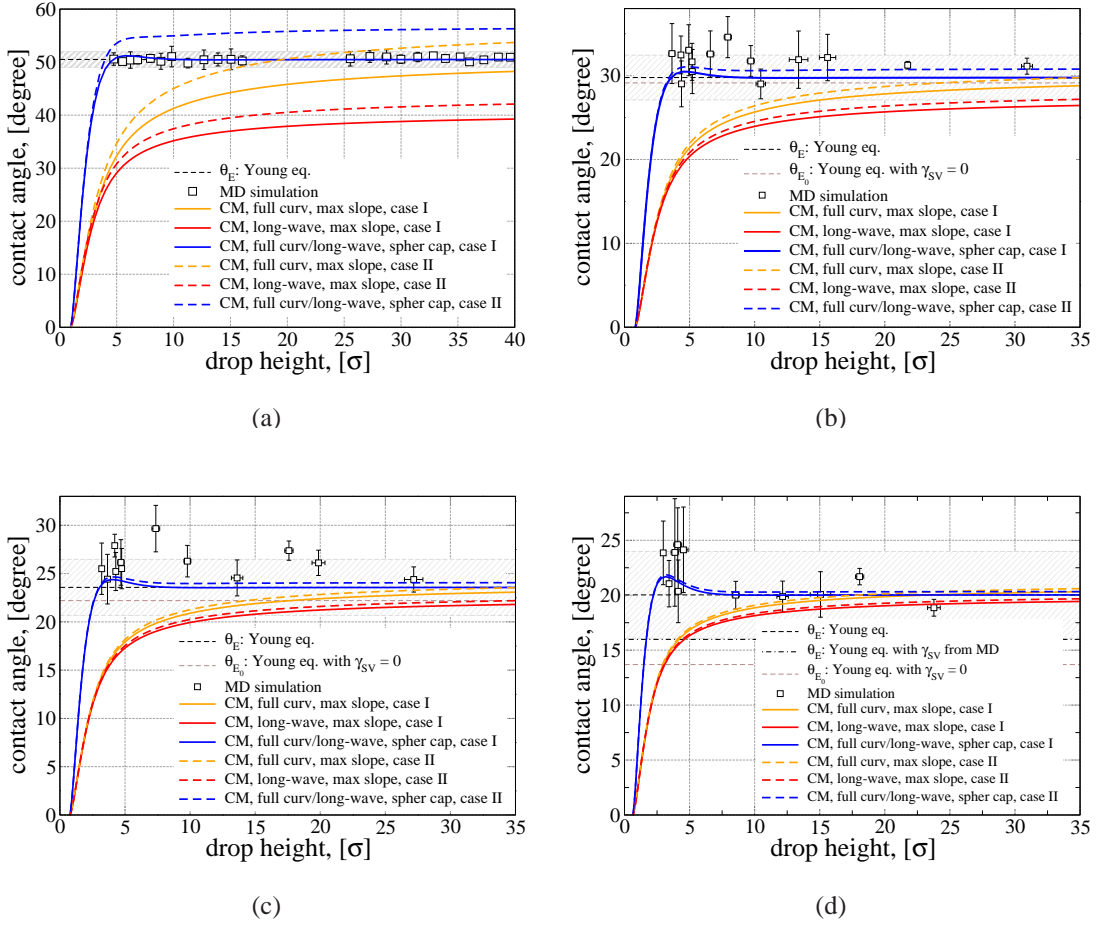

  \subfloat[]{\label{fig:ca_075}\includegraphics[width=0.42\hsize]{Fig11a_angle_height_var_eps_h0_v2_075.eps}}\hspace{0.5cm}
  \subfloat[]{\label{fig:ca_080}\includegraphics[width=0.42\hsize]{Fig11b_angle_height_var_eps_h0_v7_080.eps}}\vspace*{0.3cm}

  \subfloat[]{\label{fig:ca_081}\includegraphics[width=0.42\hsize]{Fig11c_angle_height_var_eps_h0_v8_081.eps}}\hspace{0.5cm}
  \subfloat[]{\label{fig:ca_082}\includegraphics[width=0.42\hsize]{Fig11d_angle_height_var_eps_h0_v9_082.eps}}
  \caption{(color online) Contact angles $\theta$ of droplets of
    different sizes as a function of droplet height. Panels (a), (b),
    (c) and (d) give results at solid-liquid interaction strengths of
    $\epsilon_\textrm{s} = 0.75 \epsilon$, $\epsilon_\textrm{s} = 0.80 \epsilon$, $\epsilon_\textrm{s} = 0.81
    \epsilon$ and $\epsilon_\textrm{s} = 0.82 \epsilon$,
    respectively. Square symbols correspond to the contact angle
    measured in MD using a spherical cap approximation of the droplet
    profile. Dotted and dashed thin horizontal lines correspond to the
    values $\theta_\mathrm{E_{0}}$ and $\theta_\textrm{E}$ obtained from the
    Young equation with and without accounting for the measured
    solid-vapor interface tension $\gamma_\textrm{SV}$,
    respectively. Hashed zones show the standard deviation of
    $\theta_\textrm{E}$. Panel (d) shows additionally as a dot-dashed
    horizontal line the value of $\theta_\mathrm{E}$ as extracted from the
    meniscus geometry. Case I and II refer to
    usage of only $\gamma$ or the full $\gamma+g(h)$ as prefactor of
    the curvature, respectively. 
    The thick solid and dashed curves (orange, red and blue online) in panels (a) to (d) give the mesoscopic steepest slope contact angle $\theta_\mathrm{mes}$ obtained from the continuum model with full and long-wave curvature [Eq.~(\ref{eq:el}) with Eqs.~(\ref{eq:pi-full}) or (\ref{eq:pi-longwave})] and the spherical cap contact angle
    $\theta_\mathrm{sph}$, respectively. In the last case, full and long-wave curvatures give the same result (see main text for details).}
  \label{fig:ca_drop_size}
\end{figure}

The resulting contact angles for drops of various sizes are presented
for different $\epsilon_\mathrm{s}$ as open square symbols in
Fig.~\ref{fig:ca_drop_size}.  Overall, they agree well with the
prediction of Eq.~(\ref{eq:young}) that is given as horizontal dashed
black line (with the standard deviation indicated as a grey hatched
region). Corresponding results for the contact angle obtained
from the continuum model, employing the long-wave approximation for
the curvature, Eq.~(\ref{eq:pi-longwave}), and with the full
curvature, Eq.~(\ref{eq:pi-full}), are given as well.  The results
for both are shown as solid (case I) and dashed (case II) lines of
different colors depending on the angle shown ($\theta_\mathrm{mes}$
or $\theta_\mathrm{sph}$) and the curvature used. Note that both
curvature models result in identical results for
$\theta_\mathrm{sph}$ because $\partial_x h=0$ at the apex of the
drop. This is not the case for $\theta_\mathrm{mes}$.  Case I and II
refer to the usage of only $\gamma$ or the full $\gamma+g(h)$ as
prefactor of curvature, respectively
[cf.~Eqs.~(\ref{eq:pi-longwave}) and~(\ref{eq:pi-full})].

The angle $\theta_\mathrm{sph}$ obtained in the continuum approach
agrees well with the result of the MD simulations.  This is
particularly true for case I (only $\gamma$ as prefactor of
curvature) where $\theta_\mathrm{sph}$ converges for large drops to
the value obtained with the Young-Laplace equation.  The deviations
of case II from case I are small over the entire thickness range for
$\epsilon_\textrm{s} = 0.82 \epsilon$, $\epsilon_\textrm{s} = 0.81
\epsilon$ and $\epsilon_\textrm{s} = 0.80 \epsilon$, but rather
large for $\epsilon_\textrm{s} = 0.75 \epsilon$. Note, that
$\theta_\mathrm{mes}$ does not agree well with the macroscopic angle
obtained in the MD simulations. In long-wave approximation it is
always at least some percent smaller than $\theta_\mathrm{sph}$ (more
so for small droplets). The angle $\theta_\mathrm{mes}$ obtained with
the full curvature differs less from $\theta_\mathrm{sph}$, the
difference becomes less than one percent for large drops. For
both curvature models, $\theta_\mathrm{mes}$ always decreases
monotonically with decreasing drop size. All these statements
apply for the respective relation between the various curves in case
I equally as in case II. The various angles calculated in case I are
always slightly below the ones obtained in case II.

\clearpage
Inspecting Fig.~\ref{fig:ca_drop_size}, one notes a number of further details that warrant to be highlighted: (i) A common feature of the particle-based model for $\epsilon_\textrm{s}\ge0.80\epsilon$, shown in Figs.~\ref{fig:ca_080} - \ref{fig:ca_082}, is the overshooting of the values of contact angles at thicknesses $h \approx 3 - 7 \sigma$. This effect can also be observed in the spherical cap contact angle obtained from the continuum models. It indicates that the product of drop height $h_\mathrm{max}$ and curvature at the drop apex $\kappa_\mathrm{max}$ is not a constant any more, instead $|h_\mathrm{max}\kappa_\mathrm{max}|$ first increases with increasing volume (before decreasing again).
(ii) Another detail one notices is the importance of the solid-vapor interface tension, $\gamma_\textrm{SV}$, measured in Sec~\ref{sec:pass:sub-vap}. At $\epsilon_\textrm{s} = 0.75 \epsilon$ it equals zero and at $\epsilon_\textrm{s} = 0.80 \epsilon$ the macroscopic contact angles are almost the same if one neglects $\gamma_\textrm{SV}$ or properly accounts for it (cf.\ the dotted and dashed horizontal lines in Fig.~\ref{fig:ca_080}, respectively). However, the difference between the two approaches becomes increasingly important with increasing $\epsilon_\textrm{s}$, i.e.\ decreasing contact angle (Figs.~\ref{fig:ca_081} and \ref{fig:ca_082}). Taking a non-zero $\gamma_\textrm{SV}$ into account becomes crucial close to the wetting transition. There, for rather small values of the contact angle (about $15-20^\mathrm{o}$) the difference is of the order of $20-40\%$ and accounts for $2-6^\mathrm{o}$. The difference can lead to an incorrect prediction of the contact angle if one assumes $\gamma_\textrm{SV} = 
0$ in the particle-based model. 

Finally, we note that the error bars of the contact angles $\theta$ measured in MD using a spherical cap approximation of the droplet profile (open squares in Figs.~\ref{fig:ca_075} to \ref{fig:ca_082}) are quite large. They increase with decreasing contact angle even in absolute terms.  Several possible explanations exist for this behavior: (i) In the vicinity of the wetting transition, there are strong capillary waves on the surface of the droplet (particularly close to the three phase contact line) \cite{Abraham1993}. (ii) The crossing criterion we apply to define the profile of the drops $(\rho_{0}+\rho_V)/2$ is not a unique choice. There are other possibilities to define the local interface position based, e.g., on 10-90\% or 20-80\% rules  (cf.~\cite{AllenTild89, JR_CMcC_PC_2003, AH_JH_1994}).

Next, we compare the drop profiles as obtained from the particle-based
model and the continuum description.  For the case of a rather small
contact angle, $\epsilon_\mathrm{s}=0.82\epsilon$,
Fig.~\ref{fig:drop_prof_comp082} gives results for a very small
droplet of $h_\mathrm{max}=4.046\sigma$ and a larger one with
$h_\mathrm{max}=12.181\sigma$. The layering effects of the
particle-based model are rather independent of droplet
size. Obviously, the layering of the particle-based model is not
captured by the continuum model, however, its predictions go smoothly
through the steps of the profile and always lay between the lateral
end points of the steps. At the center of the drop, the spherical-cap
fit to the particle-based model and the continuum results, obtained
with Eq.~(\ref{eq:el}) with the full curvature
(Eq.~(\ref{eq:pi-full})) as well as in long-wave approximation
(Eq.~(\ref{eq:pi-longwave})), nicely agree with each other. As cases I
and II can not be distinguished by eye alone we have only included
case I.

Differences between long-wave and full curvature and the results of
the particle-based model are only visible in the contact line
region. There, the spherical cap is not a good fit to the
particle-based model. The two continuum models nearly coincide,
implying that the long-wave approximation for static droplets is still
very good for contact angles around 20$^\mathrm{o}$.  In the contact
line region, they seem to represent a better approximation to the
particle-based model than the spherical cap. One should actually
expect this, as the continuum models incorporate the Derjaguin pressure as measured in the particle-based model. One may conclude that within its limitations the continuum model describes the profiles rather well if it incorporates the interface tensions and Derjaguin pressure from particle-based model. 

\begin{figure}[t]
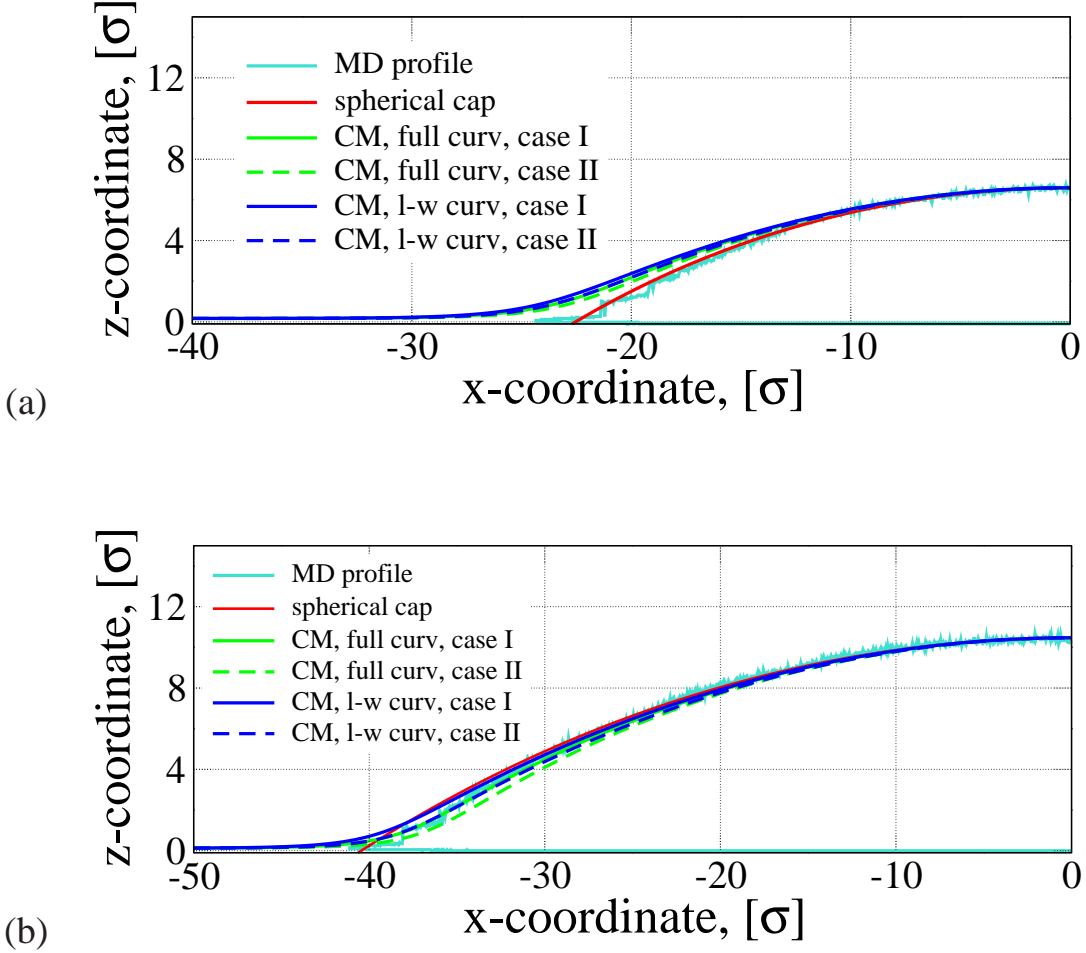

  {\large (a)}\hspace*{0.5cm} \includegraphics[width=0.8\hsize]{Fig12a_eps_080_n0450_f10_comparison_hp.eps}\vspace*{1.4cm}
  
  {\large (b)}\hspace*{0.5cm} \includegraphics[width=0.8\hsize]{Fig12b_eps_080_n1200_f27_comparison_hp.eps}
%
  \caption{(color online) Droplet profiles obtained in MD simulations and with continuum models are compared for the case
    $\epsilon_\mathrm{s}=0.80\epsilon$, apex heights (a) $h_\mathrm{max}=6.594 \sigma$ and (b) $h_\mathrm{max}=10.469 \sigma$. The solid curves (light blue online) give the liquid-vapor interface as obtained in the MD simulation, while the gray solid curves (red online) give the corresponding spherical cap fit. Results of the continuum model (CM) Eq.~(\ref{eq:el}) with the full curvature (Eq.~(\ref{eq:pi-full}) - green curves) and in long-wave approximation (Eq.~(\ref{eq:pi-longwave}) - dark blue curves) are shown for cases I and II as solid and dashed lines, respectively. For details see main text.}
  \label{fig:drop_prof_comp080}
\end{figure}
The situation differs for larger contact angles as obtained for $\epsilon_\mathrm{s}=0.80\epsilon$ and shown in Fig.~\ref{fig:drop_prof_comp080}: (i) The deviation from the spherical-cap approximation is more significant than for the smaller contact angle and (ii) the continuum model fails to describe the simulation data for the smaller droplet size. The difference between the predictions of the different versions of the continuum description is small compared to the deviation between the continuum models and the particle-based model. Therefore, the reason of the discrepancy is not rooted in the different approximations of the curvature. 

We note that interface fluctuations in a small droplet are strongly suppressed. Therefore, one should rather use the bare interface potential (i.e., interface potential without accounting of capillary waves that could be obtained from a thin film with very reduced lateral dimensions) than the one deduced from a laterally extended film. Since the bare interface potential has a smaller range than the renormalized one \cite{Lipowsky87} that accounts for thermal fluctuations of the liquid-vapor interface (i.e., capillary waves), we expect the profile of a small droplet to be better approximated by a spherical-cap shape than that of a large one, which is indeed consistent with the simulation data.

Out of the same reason, the predictions of the continuum model are more accurate for the larger drop than for the smaller one because it uses the renormalized interface potential as input. This rational explains why the predictions of the continuum model systematically deviate from the results of the particle-based model for small droplet size. For the large droplet, in contrast, the continuum model succeeds in describing the deviations from the spherical cap shape, which is larger for small contact angles. The profile of the particle-based model lays right in the middle of the predictions of the continuum models. The one that fits best is the case I with full curvature. Therefore, we conclude that even for contact angles of about 30$^\mathrm{o}$ all models agree fairly well with the particle-based simulations provided the appropriate interface potential is used.

\begin{figure}[t]
\includegraphics[width=0.8\hsize]{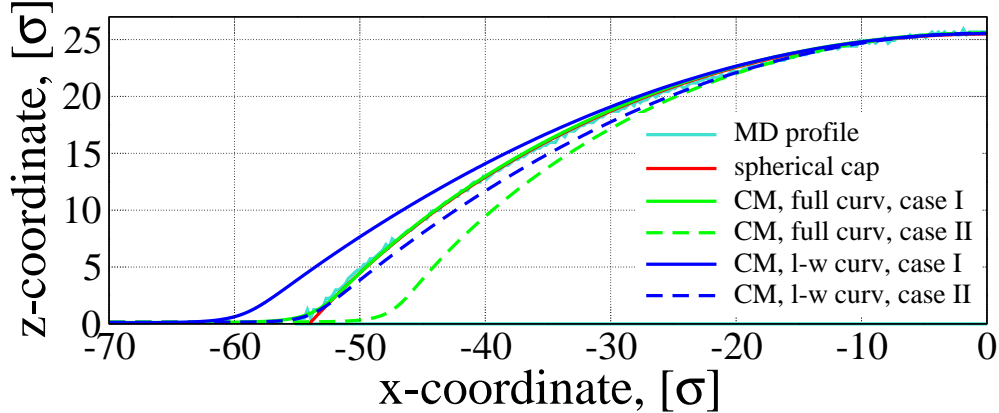}
  \caption{(color online) Droplet profiles obtained in MD simulations and with continuum models are compared for the case
     $\epsilon_\mathrm{s}=0.75\epsilon$ and apex height $h_\mathrm{max}=25.494 \sigma$. The solid curve (light blue online) gives the liquid-vapor interface as obtained in the MD simulation, while the gray solid curve (red online) gives the corresponding spherical cap fit. Results of the continuum model (CM) Eq.~(\ref{eq:el}) with the full curvature (Eq.~(\ref{eq:pi-full}) - green curves) and in long-wave approximation (Eq.~(\ref{eq:pi-longwave}) - dark blue curves) are shown for cases I and II as solid and dashed lines, respectively. For details see main text.}
  \label{fig:drop_prof_comp075}
\end{figure}
Finally, we compare the profiles with a rather large contact angle as obtained for $\epsilon_\mathrm{s}=0.75\epsilon$ and shown in Fig.~\ref{fig:drop_prof_comp075}. For comparison we use a large droplet with $h_\mathrm{max}=25.494 \sigma$. The difference between the various versions of the continuum models is clearly seen not only at the contact line but over the entire droplet profile. The best agreement with the particle-based model is achieved for case I with full curvature; all other versions differ more significantly.  Therefore, we conclude that for contact angles of about 50$^\mathrm{o}$ only the model with full curvature agrees well with the particle-based model, while the long-wave approximation is not valid anymore. It is not advisable to apply at $\theta_\mathrm{E}=50^\mathrm{o}$ where it predicts a contact angle $\theta_\mathrm{mes}$ that is 20\% lower.

\section{Conclusion and outlook} 
\label{sec:conc}
%
The equilibrium properties of polymer droplets have been studied by
Molecular Dynamics simulation of a coarse-grained particle-based model
and a continuum description in terms of an effective interface
Hamiltonian. We have devised a simple method to compute the interface
potential for laterally corrugated substrates, which is based on the
anisotropy of the pressure inside the film. This general computational
strategy can be applied to dense liquids of large macromolecules and
can be implemented in standard Molecular Dynamics programs. Using the
so-determined interface tensions and the interface potential in the
continuum model, we find quantitative agreement between both
descriptions if (i) the full curvature is used in the continuum model
for large contact angles and (ii) the size of the drop is larger than
the lateral correlation length, $\xi_{\|}$, of interface
fluctuations. We also find that for contact angles up to
 about 30 degree the long-wave approximation that is normally used 
in thin film models describes the droplet shapes even quantitatively
quite well.

These results demonstrate that the tensions and the interface
potential capture the relevant information that needs to be passed on
to a continuum model to describe the equilibrium shape of droplets,
including the deviations from the spherical cap shape in the vicinity
of the three-phase contact line. This is an excellent starting point
for comparing the dynamics of droplets driven by external forces,
which we will pursue in the future.

\section*{Acknowledgment} 
\label{sec:ack}
%
The authors thank F. L\'{e}onforte and A. Galuschko for fruitful and inspiring discussions and the GWDG computing center at G\"{o}ttingen, J\"{u}lich Supercomputing Centre (JSC) and Computing Centre in Hannover (HLRN) for the computational resources. We appreciate Stephan Kramer for the access to the GPU facilities of the CUDA Teaching Centre (CTC) at G\"{o}ttingen. This work was supported by the European Union under grant PITN-GA-2008-214919 (MULTIFLOW).

%

\end{document}